\documentclass[preprint]{aastex}
\usepackage{natbib}
\usepackage{graphics}
\usepackage{color}
\usepackage{lscape}
\usepackage{rotating}
\usepackage{footnote}
\usepackage{datetime}
\usepackage{amsmath}
\usepackage{mathrsfs}
\usepackage[normalem]{ulem}
\usepackage{CJKutf8}

\def\msun{\,  {\rm M_\odot}}
\def\cm-3{\,{\rm cm^{-3}}}
\def\erg{\, {\rm erg}}
\def\kpc-3{\,{\rm kpc^{-3}}}
\def\myr-1{\,{\rm Myr^{-1}}}
\def\pc{\,{\rm pc}}
\def\kpc{\,{\rm kpc}}
\def\yr{\,{\rm year}}
\def\fvh{$f_{\rm{V,hot}}$ }
\def\eth{$E_{\rm{th}}$ }
\def\ek{$E_k$ }

\begin{document}

\title{Supernova Feedback and the Hot Gas Filling Fraction of the Interstellar Medium
}

\author{Miao Li (\begin{CJK}{UTF8}{gbsn}李邈\end{CJK})\altaffilmark{1}, Jeremiah P. Ostriker \altaffilmark{1,2}, Renyue Cen\altaffilmark{2}, Greg L. Bryan\altaffilmark{1}, Thorsten Naab\altaffilmark{3}}

\altaffiltext{1}{Department of Astronomy, Columbia University, 550 W120th Street, New York, NY 10027, USA, miao@astro.columbia.edu }

\altaffiltext{2}{Department of Astrophysical Sciences, Princeton University, Peyton Hall, Princeton, NJ 08544, USA}
\altaffiltext{3}{Max-Planck Institut für Astrophysik, Karl-Schwarzschild-Str. 1, D-85741 Garching, Germany}

\begin{abstract}
Supernovae (SN), the most energetic stellar feedback mechanism, are crucial for regulating the interstellar medium (ISM) and launching galactic winds. We explore how supernova remnants (SNRs) create a multiphase medium by performing 3D hydrodynamical simulations at various SN rates, $S$, and ISM average densities, $\bar{n}$. The evolution of a SNR in a self-consistently generated three-phase ISM is qualitatively different from that in a uniform or a two-phase warm/cold medium. By travelling faster and further in the low-density hot phase,  the domain of a SNR increases by $>10^{2.5}$. Varying $\bar{n}$ and $S$, we find that a steady state can only be achieved when the hot gas volume fraction $f_{\rm{V,hot}}\lesssim 0.6 \pm 0.1 $. Above that level, overlapping SNRs render connecting topology of the hot gas, and the ISM is subjected to thermal runaway. Photoelectric heating (PEH) has a surprisingly strong impact on $f_{\rm{V,hot}}$. For $\bar{n}\gtrsim 3 \cm-3 $, a reasonable PEH rate is able to suppress the thermal runaway. Overall, we determine the critical SN rate for the onset of thermal runaway to be
$S_{\rm{crit}} =  200 (\bar{n}/1\cm-3)^k (E_{\rm{SN}}/10^{51}\erg)^{-1} \kpc^{-3} \myr-1$, where $k = (1.2,2.7)$ for $\bar{n} \leq 1$ and $> 1\cm-3 $, respectively. We present a fitting formula of the ISM pressure $P(\bar{n}$, $S$), which can be used as an effective equation of state in cosmological simulations. Despite the 5 orders of magnitude span of $(\bar{n},S)$, the average Mach number varies little: $\mathcal{M} \approx \ 0.5\pm 0.2, \ 1.2\pm 0.3,\ 2.3\pm 0.9$ for the hot, warm and cold phases, respectively.

\end{abstract}

\section{Introduction}
\label{intro}

The ISM permeates throughout galaxies, absorbs and scatters the light from astronomical objects, gives birth to stars, and provides material for galactic outflows. Knowledge about the ISM determines our ability to interpret astronomical data and to apprehend many fundamental astrophysical processes. The ISM is multiphase in nature. The hot component ($T \gtrsim 10^6$ K) contains little mass, but occupies roughly half the volume in the solar neighbourhood \citep{snowden98}, and may have the potential to generate galactic winds \citep[e.g.][]{li13}. How the multiphase structure forms and evolves, however, is still poorly understood. 

SN play a key role in shaping the ISM by injecting energy, momentum, mass and metals. They are the most energetic of the stellar feedback processes (radiation from stars yield more energy in total, but the efficiency of its coupling to gas is lower). SN explosions drive blast waves into the ISM, create the hot phase, generate turbulence and accelerate cosmic rays. 
A better understanding of SN feedback is therefore crucial to study the evolution of the multiphase ISM.

Recent progress in observation and simulation makes it an ideal time to revisit the problem of SN feedback and the multiphase ISM. New observational techniques have started to resolve the individual star formation sites in external galaxies, allowing us to study stellar feedback in different environment at sub-kpc scale. A correlation between the gas energy and star formation $30-40$ Myr ago, which is the typical lifetime of the progenitors of core collapse SN, implies the importance of SN feedback \citep{stilp13}. Studies of 30 Doradus, a well-resolved star forming site in the Large Magellanic Cloud, offer precious information on stellar feedback in detail \citep [e.g.][]{chu94, pellegrini10, lopez11}. Light echoes from SN themselves reveal the 3D structure of the surrounding ISM \citep{xu95,kim08}. The velocity dispersion of H$\alpha$ emission may have a mild correlation with the local star formation rate at both low and high redshifts \citep {green10, green14, genzel11,arribas14, moiseev15}. 

Ab initio cosmological simulations find that feedback is central in reproducing reasonable properties of galaxies. Simulations lacking feedback produce galaxies inconsistent with observations: the masses are too big, and baryons settle to the center and form big bulges, rather than producing extended thin disks \citep{navarro91,navarro94,abadi03};  galactic winds, which are observed ubiquitously in the star-forming galaxies \citep[e.g.][]{heckman00,steidel04,steidel10, strickland09,banerji11, genzel11,bouche12,martin12}, are missing. By implementing ``feedback" the above problems are largely mitigated, since feedback heats and ejects gas, preventing gas from turning into stars, launching galactic scale winds and fountains, which reduce the mass of the bulge and transport gas into outer parts of galaxies \citep{sommer-larsen99, scannapieco12,hummels12,hopkins12,stinson13, hopkins14,ubler14}. However, due to the limited resolution, cosmological simulations can only implement SN feedback in a sub-grid manner, and most of the current simulations use ad hoc models which involve turning off standard physics such as cooling or gas dynamics. As a result, the predictive power of such simulations is seriously limited. A more physically-based model of SN feedback is essential to improve the current galaxy simulations, and to achieve this, a better understanding of the interplay between SN and the multiphase ISM is indispensable. 

The evolution of a SNR in a uniform medium is well-studied by many authors. Early 1D simulations \citep[e.g.][]{chevalier74, cioffi88} showed that the evolution is characterized by four stages: free expansion, Sedov-Taylor, ``pressure-driven snowplow'' and momentum-conserving. \cite{thornton98} revealed the feedback at later stages is dominated by kinetic energy while thermal feedback is negligible due to rapid cooling. Taking into account the kinetic feedback may have big implications for galaxy formation \citep[][N\'u\~nez et al. 2015, in prep]{simpson15}. 

SNRs evolve in a qualitatively different way when exploding in a clumpy medium. Pioneer work by \cite{cowie81} studied the remnant evolution and energy budget in a three phase medium with most volume hot, and found that the blast wave travels much faster and further in the low-density, hot medium, and the loss of explosion energy happens later compared to the case when the medium is uniform. Recent 3D simulations by \cite{martizzi15} and \cite{kim15} confirmed that the evolution and energy feedback of a SNR in a clumpy medium is quite different from the uniform case, but the terminal radial momentum is affected little by the inhomogeneity of the ISM. \cite{walch_naab15} found similar results for molecular clouds, and the pre-SN ionization can slightly enhance the final momentum per SN.

Recently works on SN feedback have mainly considered a relatively dense medium and neglect the third phase of the ISM -- the tenuous hot gas. However, a medium with a significant fraction of volume hot exists widely, e.g. at the disk-halo interface and the galactic winds launching sites. Moreover, the hot phase being the most tenuous transports energy/momentum much faster (speed of sound/blast wave $\propto \rho^{-0.5}$) and cools very inefficiently (cooling time $\propto \rho^{-1}$). It is therefore very interesting to investigate the SN feedback in a three-phase ISM. 

The multiphase ISM adopted in the studies of SN evolution has been largely set up by hand. However, we should note that while SNRs are affected by the lumpiness of the ISM, they are also the major player in shaping the overall properties of the gas. Thus it is very important to study SN feedback self-consistently.
\cite{mckee77} proposed the underlying theory of the three-phase ISM regulated by SN. Their basic ideas were the following:

(i) If the SN rate is sufficient to render the porosity of the ISM $\sim 1$, then the ISM is inevitably multi-phase with the majority of volume being hot;

(ii) All three phases are in approximate pressure equilibrium \citep{spitzer56};

(iii) Different phases are in dynamical equilibrium 
in terms of mass and energy: SN keep creating hot bubbles and compressing diffuse gas into dense clouds, while clouds are evaporated by the hot medium; energy input from SN balances dissipation in all phases. 

These principles yield a complete set of equations, and the solutions determine the pressure and mass/volume quota for each phase, given the mean gas density and SN rate. Their results were in rough agreement with observations of the ISM in the solar neighbourhood. Despite some imperfectness of the theory (for example, underestimating of the mass of warm gas, not considering the clustering of SN, etc), their main ideas are still widely accepted as the standard picture of the local ISM. Numerical simulations clearly confirm that three phases coexist in rough pressure equilibrium, with temperature around $10^2, 10^4, 10^6$ K, separately, and the hot gas contains a significant fraction of volume but little mass \citep{deavillez04, joung06,joung09, creasey13, gent13,walch14}. \cite{mckee77} applied their theory primarily to the ISM around us. We now want to extend to very different environments, where the gas density and SN rate vary significantly from that in the solar neighbourhood, such as the Galactic halo, a starburst region, etc, where the multiphase structure of the ISM can vary vastly. In particular, when the SN rate is high enough, an equilibrium state no longer exists and wind generation is inevitable \citep[e.g.][]{maclow88, scannapieco13, gatto15}. 

An aspect of SN feedback that has long been neglected is that a significant fraction of SN progenitors are the so-called ``OB runaways" \citep[e.g.][]{gies86, stone91,hoogerwerf01,dewit05,tetzlaff11}, thus SN can explode far from star formation sites. OB stars acquire this velocity when they reside in close binary systems and their companions explode as SN \citep{zwicky57, blaauw61}, and/or when they are in a crowded region of star clusters and get ejected due to dynamical interactions \citep{poveda67}. \cite{stone91} observed that $46\%$ of O stars and $10\%$ of B stars are in the ``runaway" category with $\sigma \sim 30$ km/s. This means that OB stars can travel $50-500 \pc$ before exploding as SN. SN can therefore deposit energy outside dense molecular clouds, and may go into the low density inter-spiral arm or halo. 
The spatial distribution of the pulsars independently confirms this idea: the scale height of pulsars at birth is much larger than that of the Galactic HI layer. \citep[cf.][]{narayan90}. Consequently, the SN energy is much less prone to radiative losses due to low gas density. Type Ia SN, which are associated with the older stellar population with $h \sim 300 \pc$, have similar effects. The ``runaway" OB stars together with Type Ia SN are thus very likely to contribute to the launching of galactic outflow out of proportion to their intrinsic frequency \citep{ceverino09,kimm14,walch14}, but this effect has been explored very little. The velocities of OB stars also determine the spatial distribution of the core collapse SN, which is important for the resultant ISM properties. For example, \cite{gatto15} found that random positioning of SN leaves most volume hot and thermal runaway, while SN exploding at density peaks result in little hot gas.

In this paper we explore the interaction between SN and multi-phase ISM. 
The primary questions we wish to address in this paper are:

a. Given a multiphase ISM with a significant volume hot, how does a SNR evolve and impart feedback? In particular, how does the feedback differ from that in a uniform medium?

b. Given a mean gas density $\bar{n}$ and SN rate density $S$, what is the resultant ISM, especially $f_{\rm{V,hot}}$?

We propose three sets of experiments: First, as a baseline, we evolve a single SN in a uniform medium, to compare with the previous studies and parametrize the feedback effects. 
Second, we study how a SN evolves and imparts momentum and energy in an inhomogeneous medium, and compare it with the uniform case.
Third, we allow multiple SN to shape the medium. We investigate the parameter space $(\bar{n},S)$ and analyze the resultant multiphase structure of the ISM. To be self-consistent, the multiphase ISM models in the second experiment are taken from the third.

The paper is organized as follows: In Section 2, we describe the code we use and relevant physics included; in Section 3, we present the single SN feedback in uniform and multiphase media; in Section 4, we show how multiple SN shape the multiphase ISM; we discuss our results in Section 5, and conclude in Section 6.

\section{Numerical Methods}
The numerical experiments are performed on a uniform-mesh code which uses the shock-capturing total variation diminishing (TVD) scheme \citep{Ryu93} to solve the hydrodynamics equations. The ideal gas law is applied with the adiabatic index $\gamma = 5/3$. Cooling and heating are calculated using the Grackle library \footnote{http://grackle.readthedocs.org/en/latest/} as described in \cite{smith08, smith11}. Metallicity is assumed to be solar, that is, 2\% of gas mass is in metals. Cooling and heating rates due to H, He and metals (including fine-structure lines) at a given density and temperature are pre-calculated using the Cloudy code \citep{ferland98}, which determines the ionization equilibrium solution for atom/ion species computed under an incident spectrum. In reality, the radiation field in a star formation region is very complicated and time-variable, and to be self-consistent one needs to appeal to radiative transfer. For simplicity, we do not take into account the ionizing radiation for this study. Photoelectric heating (PEH) through dust is included. We do not consider the heating of neutral gas due to cosmic rays or X-rays, since it is an order of magnitude smaller than the PEH \citep{draine11}. At each time step (determined by the Courant condition), Grackle updates the internal energy of each cell by integrating the tabulated cooling/heating rates over that time step. Self-gravity, thermal conduction, molecule formation and cooling, and magnetic fields are not included in the current simulations. 
 
We test the code against the classic Sedov-Taylor solution with cooling turned off and a resolution of 1 pc. The overall agreement is satisfactory: the shock radius agrees to within 10\% of the analytical result \citep{taylor50,sedov59}.

\section{Single SN evolution and feedback}
\subsection{In uniform media}
\label{sec:I}
In this section, we present the simulations of a SNR in a uniform medium, and analyze how a SN imparts its energy and momentum to the gas. We also compare our results with other recent numerical works. 

\subsubsection{Fiducial run: Setup and Results}
\label{simIfid}

We set up our fiducial run as follows: initially the gas is uniform with $n =1 \cm-3$ and $T = 8\times 10^3$ K. We inject $E_{\rm{SN}}=10^{51}$ erg as purely thermal energy and $10 \msun$ mass into a sphere which encloses about $40 \msun$ of the ISM. The initial sphere has a radius of 7.5 pc. Both energy and mass are evenly distributed. The spatial resolution is 0.75 pc, and the box is 225 pc on one side. Periodic boundary conditions are applied, although the SNR does not reach the boundary. Background PEH is set to be $1.4 \times 10^{-26}$ erg/s per hydrogen atom, which is similar to that in the solar neighbourhood \citep{draine11}. 

\begin{figure*}
\begin{center}
\includegraphics[width=1.1\textwidth]{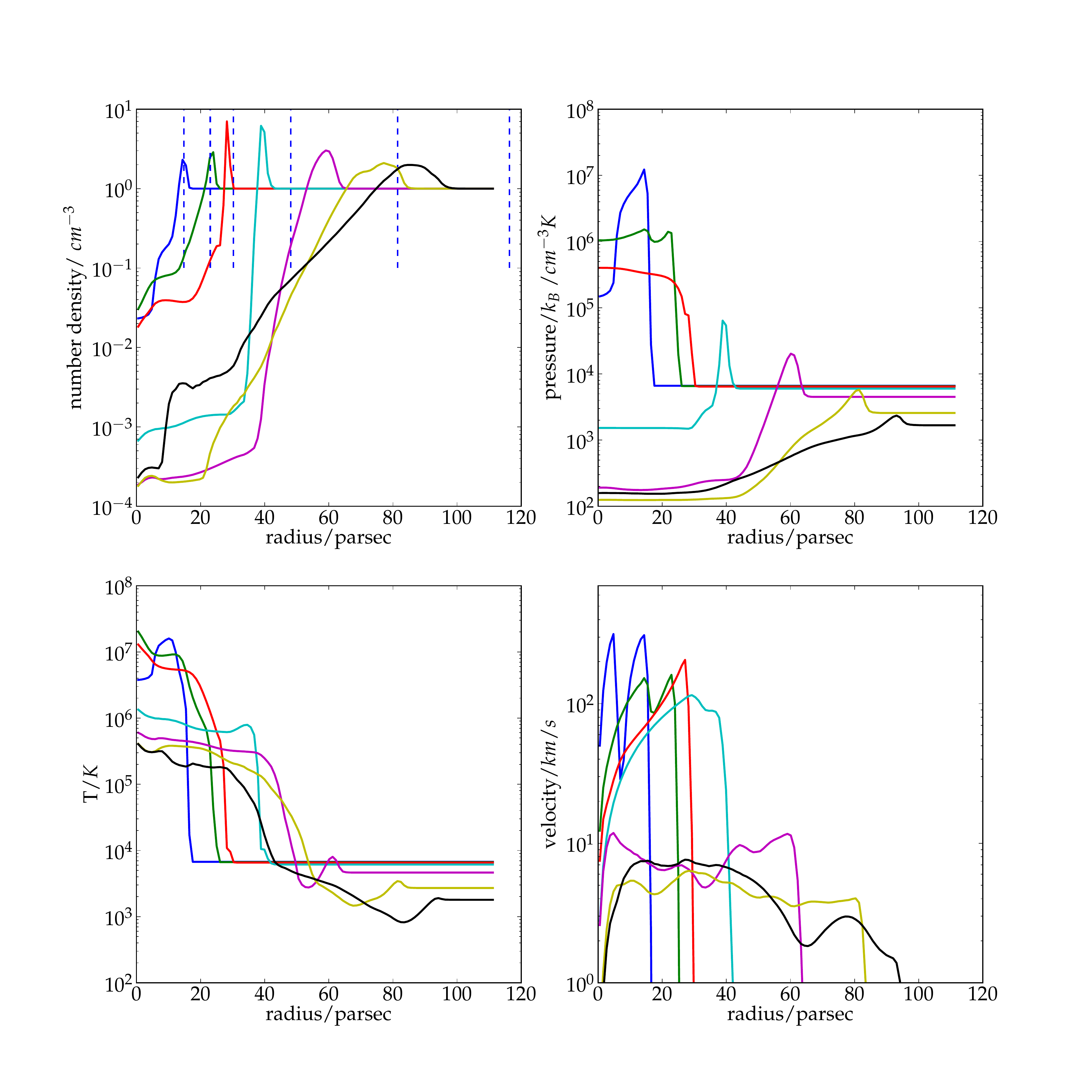}
\caption{Evolution of a SNR of initial energy $10^{51}$ erg in a uniform medium with $n=1\cm-3$. Snapshots are taken at $t = 1.5 \times 10^4$, $4.6\times 10^4$, $9\times 10^4$, $2.9\times 10^5$, $1.1\times 10^6$, $2.6\times 10^6$, $4.2\times 10^6$ years, respectively. Numerical resolution is 0.75 pc. Vertical axes show spherically-averaged quantities, and horizontal axes are radii. Dashed lines in the upper-left panel show the positions of the shock as predicted by the ST solution at the above timesteps. Pressure of the inner bubble drops precipitously once the thin shell forms, to even below that of the ISM. At later stage, the shell becomes thicker and keeps moving outward to $R>80\pc$, while the the hot bubble ($T>2\times 10^5$ K) barely reaches 42 pc in radius. Velocity is shown in absolute value.}
\label{f:single}
\end{center}
\end{figure*}

Fig. \ref{f:single} shows the physical quantities (spherically-averaged) as a function of radius. Dashed lines in the upper-left panel indicate the shock radius as predicted by the Sedov-Taylor (ST) solution for the 7 timesteps we show. The position of the shocks in our simulation coincide with theory very well in the energy conserving phase. The peak density at this stage is $\sim$ 2.5$n$, somewhat smaller than that in the ST solution, i.e. 4$n$. This is likely due to the finite resolution and the application of spherical-averaging. 
After cooling becomes important, the blast wave slows down and lags behind the ST solution. A thin, dense and cool ($T\sim 10^4$ K) shell forms, and the SNR enters the ``pressure-driven snowplow" phase. We define $R_{\rm{0.8Eth}}$ as the radial position of the shock when 20\% of the thermal energy in ST stage has been lost due to cooling. We find $R_{\rm{0.8Eth}}=22.1\pc$ at $t = 4.1 \times 10^4$ year, in good agreement with the analytical calculation by \cite{draine11}: 
\begin{equation}
{R_{\rm{cool}}} = 23.7 \pc (\frac{n}{1\cm-3})^{-0.42} (\frac{E_{\rm{SN}}}{10^{51}\erg})^{0.29},
\label{eq:rcool}
\end{equation}
\begin{equation}
t_{\rm{cool}} = 4.9\times 10^4 \yr (\frac{n}{1\cm-3})^{-0.55} (\frac{E_{\rm{SN}}}{10^{51}\erg})^{0.22}.
\label{eq:tcool}
\end{equation}
The equations are obtained by assuming the cooling rate $\Lambda \propto T^{-0.7}$ for $10^{5-7.3}$K gas. Note that in \cite{draine11}, $R_{\rm{cool}}$ and ${t_{\rm{cool}}}$ are defined somewhat differently, as the radius and time when 30\% of the SN energy has been radiated away.

While the shell has cooled and slowed down, the inner bubble is still filled with gas that is hot ($T>10^6 $ K) and fast-moving ($v>$ 100 km/s). During this stage, the density of the hot bubble drops rapidly. This is because the fast-moving gas keeps running into the cool shell where it radiates energy very efficiently, which leads to substantial mass loss from the hot bubble. 
Accompanying the mass loss is a sharp pressure decrease, to even below that of the ambient gas. The precipitous decline of density and pressure makes the central engine quickly run out of steam. Thus, after a short-lived ``pressure-driven snowplow" phase, a SNR enters the ``momentum-conserving" stage. Without a propelling force, the cool shell moves forward due to its own inertia. The pressure of the shell, which is now beyond both the ISM and the inner bubble, drives itself to become thicker. Now the thick ``shell" dominates the evolution of the SNR: while the outer edge of the shell still expands with a forward shock, the hot bubble it encompasses shrinks in size simultaneously. As shown in the temperature panel (lower-left), the radius of the hot bubble ($T>2\times 10^5$ K) barely reaches 42 pc, while most swept-up mass moves to $R>80$ pc. We define $\phi$ as the ratio between the maximum size of the hot bubble and $R_{\rm{cool}}$, i.e.
\begin{equation}
\phi \equiv R_{\rm{hot,max}}/R_{\rm{cool}} \ .
\end{equation}
We find $\phi =1.77$ for the fiducial run. The parameter $\phi$ is important for the later calculation of the hot gas filling factor \fvh (see Section \ref{sec:III}). Eventually, the velocity of the front shock drops to the sound speed of the ISM, and the temperature of the shell decreases until the PEH balances the cooling. The bubble created by SN will be filled with the ambient gas again, as seen from the increase of the density of the central bubble at the end of the simulation.

\begin{sidewaysfigure*}
\begin{center}
\includegraphics[width=1.2\textwidth]{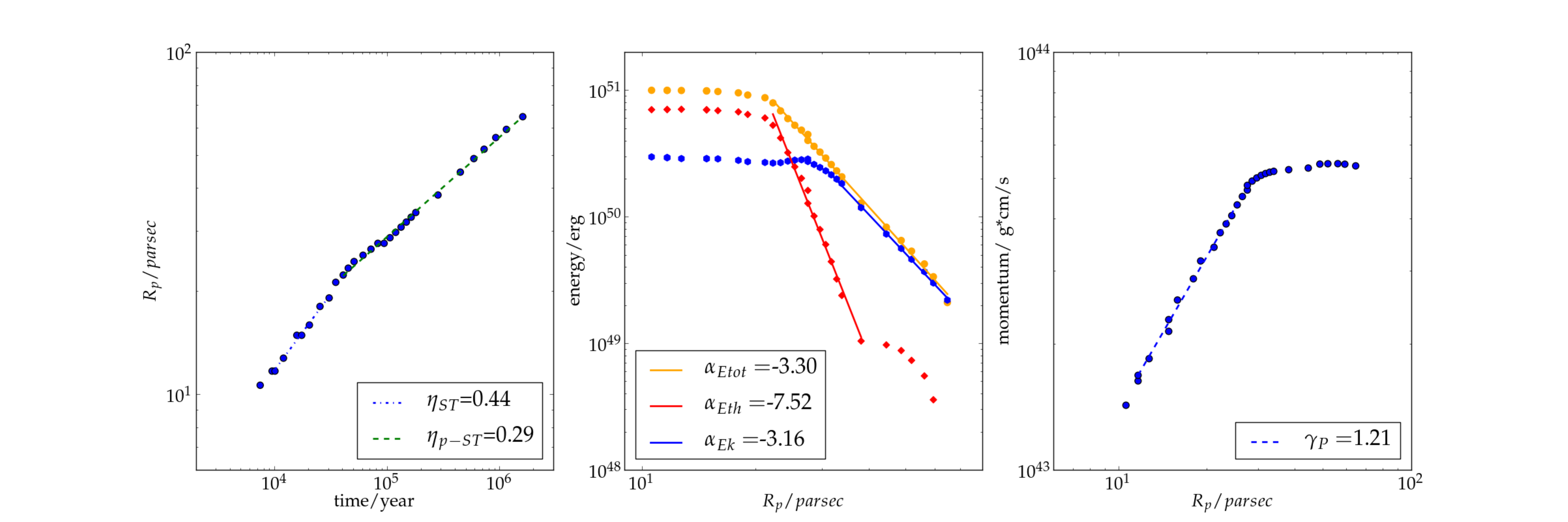}
\caption{Radial position of the blast wave $R_p$ vs time (left), energy ($E_{\rm{tot}}, E_{\rm{th}}, E_{\rm{k}}$) vs $R_p$ (middle), and momentum vs $R_p$ (right) for a $10^{51}$ erg SN in a uniform medium with $n=1\cm-3$. Power-law fits are shown with lines. The power-law indices, as defined in Eq. \ref{eq:eta_st}, \ref{eq:eta_pst},  \ref{eq:alpha_etot}-\ref{eq:gamma_st} are given in the legend. 
}
\label{f:e_r_mom}
\end{center}
\end{sidewaysfigure*}

Regarding the feedback effects, we are interested in how fast the blast wave propagates in the ISM, and how much energy and momentum it injects into the ISM as the shock reaches different radii. We present the results in Fig.\ref{f:e_r_mom}. The radius of the blast wave $R_{p}$, defined as the radius where the density peaks, follows a piecewise power-law function with time. The turnover happens at $R_p \sim R_{\rm{0.8Eth}}$. We define the power-law indices for the ST and the post-ST stage as $\eta_{ST}$ and $\eta_{\rm{p-ST}}$, respectively, i.e.
\begin{equation}
R_p \propto t^{\eta_{\rm{ST}}},\ \rm{for} \ R_p <R_{0.8Eth} ,
\label{eq:eta_st}
\end{equation}
\begin{equation}
R_p \propto t^{\eta_{\rm{p-ST}}},\ \rm{for}\ R_p  \geqslant R_{0.8Eth}.
\label{eq:eta_pst}
\end{equation}
A simple linear fit on log-log scale using the ordinary least squares method yields: $\eta_{\rm{ST}}=0.445 \pm 0.020 $ and $\eta_{\rm{p-ST}}=0.292 \pm 0.004 $. The ST solution predicts $\eta_{\rm{ST}}=0.40 $. \cite{chevalier74} found $\eta_{\rm{p-ST}} \approx 0.31$; \cite{blondin98} had a somewhat higher value 0.33.

The plot of energy vs $R_p$ (middle panel of Fig. \ref{f:e_r_mom}) indicates that the decline of kinetic energy $E_k$ happens after that of thermal energy $E_{\rm{th}}$. We use the parameter $\beta$ to characterize this delay:
\begin{equation}
\beta \equiv R_{\rm{0.8Ek}}/R_{\rm{0.8Eth}},
\end{equation}
where $R_{\rm{0.8Ek}}$ is defined similarly as $R_{\rm{0.8Eth}}$. Our simulation indicates $R_{\rm{0.8Eth}} = 22.1 \pc$, $R_{\rm{0.8Ek}}=30.5 \pc$ and thus $\beta = 1.38$. The numerical value of $\beta$ is understandable, since this process can be seen as an inelastic collision between the cool shell and the ISM, during which $E_k$ only undergoes significant loss when the sweep-up mass is similar to that of the shell (38\% increase in radius means roughly a 162\% increase in volume and thus, mass). The decline of $E_{\rm{tot}}$, $E_{\rm{th}}$ and $E_{k}$ seem to occur in a power-law fashion, and we define the corresponding indices $\alpha$ as follows:
\begin{equation}
E_{\rm{tot}} \propto R_p ^{\alpha_{\rm{Etot}}}, \ \ \rm{for}\ R_p >R_{\rm{0.8Eth}},
\label{eq:alpha_etot}
\end{equation}
\begin{equation} 
E_{\rm{th}} \propto R_p ^{\alpha_{\rm{Eth}}}, \ \ \rm{for}\ R_p >R_{\rm{0.8Eth}},
\label{eq:alpha_eth}
\end{equation}
\begin{equation}
E_{k} \propto R_p ^{\alpha_{\rm{Ek}}}, \ \ \rm{for}\ R_p >R_{\rm{0.8Ek}}.
\label{eq:alpha_ek}
\end{equation}
We find $\alpha_{\rm{Etot}}= -3.30 \pm 0.04 $, $\alpha_{\rm{Eth}}= -7.52 \pm 0.21 $,\ $\alpha_{\rm{Ek}}= -3.16 \pm 0.03 $. The loss of thermal energy is much more rapid than kinetic, indicating that at later stages of SN evolution, kinetic feedback dominates over thermal feedback \citep[e.g.][]{thornton98,walch_naab15,simpson15,kim15}. We note that $\alpha_{\rm{Ek}} \sim -3$ implies momentum $\mathcal{P}= (2mE_k)^{0.5} \propto (R_p^3E_k)^{0.5} \sim$ constant, that is, momentum conservation starts at $R_p>R_{0.8Ek} \sim 30 \pc$. This is confirmed by the $\mathcal{P}$ vs $R_p$ relation shown in the right panel of Fig. \ref{f:e_r_mom}. 

The momentum $\mathcal{P}$ shows a power-law rising with $R_p$, followed by a brief turn-over, and then reaches a plateau when $R_p>R_{\rm{0.8Ek}}$. Again, we define the power-law index $\gamma_P$ for the rising phase of the momentum as:
\begin{equation}
\mathcal{P} \propto R_p^{\gamma_{\mathcal{P}}} , \ \ \rm{for}\ R_p >R_{0.8Ek}.
\label{eq:gamma_st}
\end{equation}
We find $\gamma_\mathcal{P} = 1.21 \pm 0.03 $. For the ST phase, when $E_k$ is constant and $m\propto R_p^3$, a simple scaling relation shows $\mathcal{P} = (2mE_k)^{0.5} \propto R_p^{1.5}$. The final momentum $\mathcal{P}_{\rm{end}}$ is $5.4\times 10^{43}$ erg$\cdot$cm/s, consistent with Kim \& Ostriker (2015, hereafter KO15).

\subsubsection{Dependence of results on input parameters}
\label{sec:simI_all}

\begin{sidewaystable}
\begin{center}
 \caption{Best-fit parameters for the evolution of a SNR of $10^{51}$ erg in a uniform medium (definition of the symbols can be found in Section \ref{simIfid})}
  \begin{tabular}{@{}c@{}  @{}c@{}   @{}c@{}| c   c    c     c  c  c c  c @{}c@{} }    \hline 
   &Model Input && &&&Output &&\\
   \begin{tabular}{@{}c@{}} $n$ \\ $(\cm-3)$\end{tabular}&  \begin{tabular}{@{}c@{}} Res\\(pc) \end{tabular} & \begin{tabular}{@{}c} PEH  \\(erg$^{-1})$ \end{tabular} & $\alpha_{\rm{Etot}}$  & $\alpha_{\rm{Eth}}$ &$\alpha_{\rm{Ek}}$ & $\eta_{\rm{p-ST}}$ & $R_{\rm{0.8Eth}}$ & $R_{\rm{0.8Ek}}$ &$\beta$& $\phi$  &  \begin{tabular}{c} $\mathcal{P}_{\rm{end}}$\\ $(10^{43}$g$\cdot$cm/s) \end{tabular} \\ \hline
    0.005    &5.0  &  1.4E-26     &  -3.41 $\pm$ 0.06 &  -6.40  $\pm$ 0.13       & -2.96  $\pm$ 0.03 & 0.321  $\pm$ 0.003    &201.0  & 280.0 &  1.39 & $>$1.60 &10.4  \\ 
    ..    &10.0 &  ..    &  -3.89 $\pm$ 0.13 &  -5.79  $\pm$ 1.66      & -2.78   $\pm$ 0.14 & 0.334  $\pm$ 0.006    &210.0  & 290.0 &  1.38& $>$1.32 &11.0 \\     \hline
    0.1    &1.98  &  ..     &  -3.06 $\pm$ 0.10 &  -6.14  $\pm$ 0.23       & -2.40   $\pm$ 0.09 & 0.328  $\pm$ 0.006    &58.0  & 80.0 &  1.38& 1.64 &8.3 \\ 
    ..  &   3.96   &   &  -3.09  $\pm$ 0.22& -6.20  $\pm$ 0.44  & -2.33 $\pm$ 0.12   & 0.335   $\pm$ 0.012 &  57.8   &  84.5 &1.46 & 1.5  &8.5 \\ \hline
   1 & 0.25 &  .. & -3.20 $\pm$ 0.08 & -6.46  $\pm$ 0.14  &  -2.89  $\pm$ 0.07 &  0.298  $\pm$ 0.003 &  21.6  &29.8  &1.38 & $>$1.65  & 5.3 \\ 
  .. & 0.46 & .. & -3.16  $\pm$0.06 &  -6.49 $\pm$ 0.15& -3.14 $\pm$ 0.03& 0.290 $\pm$ 0.004   &  21.4 & 29.9& 1.40& $>$1.69 &5.3\\
  .. & 0.75 & ..&  -3.30 $\pm$ 0.04 & -7.52 $\pm$ 0.21  &-3.16 $\pm$ 0.03 & 0.292 $\pm$ 0.004 & 21.8 &  30.1& 1.38 & 1.77 & 5.3\\
  .. &1.5  & .. & -3.03  $\pm$  0.07&-7.64 $\pm$0.44 & -3.18 $\pm$ 0.08 & 0.293  $\pm$ 0.006 &  21.6  &29.4  &1.36&1.58 &5.3  \\  \hline
   10  & 0.56  & ..   &  -2.95 $\pm$0.09  &  -7.33  $\pm$ 0.43  & -3.00 $\pm$ 0.09   & 0.284   $\pm$ 0.005 &   7.9& 11.4 &1.44 &1.67 &3.8    \\
    ..     &    ..     &  1.4E-25  &  -3.10  $\pm$ 0.10& -7.43  $\pm$0.45    &-3.10 $\pm$0.08  & 0.277  $\pm$ 0.005 &  7.9  & 11.4 & 1.44 & 1.67 &3.8  \\
   ..  & ..     &  1.4E-23   & -3.35 $\pm$ 0.10  & -10.01  $\pm$ 1.03  & -3.20 $\pm$ 0.07  &   0.294  $\pm$0.005 & 8.3& 11.4 & 1.37 & 1.67& 3.7\\ \hline
  \begin{tabular}{@{}c@{}} 1 \\(KO15)\end{tabular} &   0.25   & 1E-26  & -   &  -5.69\footnotemark[1]  & -3.16\footnotemark[1]  &  0.29 & -&- &-& - & 5.9 \\\hline
  \end{tabular} 
       \footnotetext{$^1$measured from their Fig 1}

\end{center}
   \label{t:simI}
\end{sidewaystable}

The above scaling relations, especially the indices $\alpha$, $\beta$, $\eta$ and numerical values of $\mathcal{P}_{\rm{end}}$, are very important for SN feedback models. We want to test how the results change as we vary the numerical resolution, as well as with other model inputs, since in real galaxies, SN explode in various environments with different ISM density, PEH rate, metallicity, clumpiness, etc. In this subsection we study the effect of gas density, PEH rate and numerical resolution, and leave the task of exploring ISM clumpiness to later sections.

In Table 1 we listed the models and the outputs. For $n=1\cm-3$, we run three additional simulations with resolutions of $0.25\pc$, $0.46\pc$ and $1.5\pc$, respectively. The output parameters show variance within 10\%, and there seems no systematic change of the outputs as we improve the resolution by a factor of 6, with the exception that $\alpha_{\rm{Eth}}$ seems somewhat higher in lower resolution runs. For the two highest-resolution runs, we have only lower limits on $\phi$ due to the limitation on the box size.

Comparing four runs with different gas densities, i.e. $n=0.005$, 0.1, 1.0 and 10.0 $\cm-3$, we find that a denser ISM shows somewhat more rapid decrease of $E_{\rm{th}}$ with $R_p$, due to the more efficient cooling. But overall the difference is pretty small: decreasing $n$ from $10$ to $0.1 \cm-3$ makes $\alpha_{\rm{Eth}}$ drop by only about 20\%. KO15 also mentioned that the evolution is basically identical for different values of $n$ when the radius and time are scaled relative to their values at shell formation. For the $n=0.1$ and $0.005\cm-3$ models, $\eta_{\rm{p-ST}}$ (0.33) is larger than in higher density runs, which usually have $0.28-0.30$. $\phi$, which quantifies the volume of hot gas created by one SN, is in the range of 1.5-1.8.

$\mathcal{P}_{\rm{end}}$ is very insensitive to $n$. As presented in KO15, this can be understood using a simple analytical argument. Since 
\begin{equation}
\mathcal{P}_{\rm{end}} = (2mE_k)^{0.5} = [8\pi /3\  nm_H (\beta R_{\rm{cool}})^3 E_k ]^{0.5}, 
\label{eq:P_end}
\end{equation}
where $E_k$ is simply a constant before momentum-conserving phase. Combining Eq. \ref{eq:rcool} and \ref{eq:P_end} yields
\begin{equation}
\mathcal{P}_{\rm{end}} \sim 5\times 10^{43} \rm{g\cdot cm/s}\  (\frac{n}{1\cm-3})^{-0.13}(\frac{E_{\rm{SN}}}{10^{51}\erg})^{0.94}(\frac{\beta}{1.4})^{1.5}.
\end{equation}
In other words, in a uniform medium, SN (assuming a fixed total energy) inject an almost constant amount of momentum, independent of the density of the ISM. Our results agree with the analytical calculation fairly well. $\alpha_{\rm{Ek}}$ is somewhat larger than -3 for $n=0.1 \cm-3$, which implies that  momentum is not quite conserved at this stage, but is still slowly increasing. In this case the hot inner bubble has some pressure left to push the shell. $\mathcal{P}_{\rm{end}}$ is measured when the post-shock velocity drops to the sound speed of the ambient gas, 10 km/s.

We test how the PEH rate affects the result for $n=10 \cm-3$. From the data, we find there is no significant difference in the output parameters. One may notice that when we adopt a PEH rate 1000 times the MW value, $\alpha_{\rm{Eth}}$ is somewhat smaller, which may seem unintuitive. This is because when we calculate $E_{\rm{th}}$ of the blast wave, we deduct the ISM energy (calculated by taking a sample outside the shocked region) from the total thermal energy enclosed in the blast wave domain. The high PEH rate keeps the ISM at $\sim 10^4$ K, while for the MW value, heating cannot balance cooling and the temperature of the ISM decreases with time. This makes the remaining $E_{\rm{th}}$ inside the blast wave smaller for higher PEH rates.

Compared with KO15, where they use a resolution of $0.25$ pc for $n =  1\cm-3$, we find very good agreement for $\alpha_{\rm{Ek}}$, $\eta_{\rm{p-ST}}$ and $\mathcal{P}_{\rm{end}}$. They have a somewhat smaller $\alpha_{\rm{Eth}} \sim -5.7$, which can be due to the code and/or cooling curve we each adopt (they approximate the cooling curve by piece-wise power-law functions).  

In short, the evolution and feedback of a SNR in a uniform medium, as a function of time or $R_{p}$, seem to be well characterized by (piece-wise) power-law functions. Overall, the dimensionless numbers $\alpha$, $\beta$, $\eta_{p-St}$ and $\phi$ vary little over a large dynamic range of $n$.

\subsection{In hot-dominated multiphase media} 
\label{sec:II}
 
In this section, we study how a hot-dominated multiphase medium (HDMM) affects SN feedback. For a medium that is predominantly two-phase (warm/cold), one can refer to the recent works by KO15.

In principle, one needs many parameters to fully describe a multi-phase medium, for example, the size, density, shape of a single cloud; the location, number density and topology of clouds; the pressure of the inter-cloud medium; turbulence/velocity fields, etc, and to different extent, they all impact the evolution of a SNR. To exhaust the parameter space is not feasible. On the other hand, the above properties are not independent of each other, and more importantly, they are tightly correlated with SN feedback itself. (Other conditions, such as gravity, stellar outflows and radiation feedback, contribute to the ISM conditions as well, but arguably not as significantly as SN, and to consider these effects is beyond the scope of this paper.) For self-consistency, the ISM models adopted in this section are based on the results from Section \ref{sec:III}, where we explore how multiple SN collectively shapes the ISM. Other recent works use different methods to construct multiphase media. For example, KO15 establish the warm/cold media through thermal instability; \cite{martizzi15} impose a log-normal density structure; \cite{walch_naab15} use fractal molecular clouds, with some runs having pre-ionization.

\subsubsection{Model Description}
\label{sec:IIpara}

\begin{sidewaystable} 
\caption{Parameters of the multiphase ISM models in Section \ref{sec:II}}
(a) Artificial models\\
  \begin{tabular}{c | c    c    c  c  c|  c     c    c    c  c |  c  c  c }
    \hline
      & && Cold & & & && Warm& &&  & Hot  & \\
  Run  & \begin{tabular}{@{}c@{}} $R_{\rm{cl}}$\\(pc) \end{tabular} & \begin{tabular}{@{}c@{}} $n$\\$(\cm-3)$ \end{tabular} & \begin{tabular}{@{}c@{}} $P/k_B$\\($\cm-3$K) \end{tabular} & $f_{V}$ & $f_{m}$ & \begin{tabular}{@{}c@{}} $R_{\rm{cl}}$\\(pc) \end{tabular} & \begin{tabular}{@{}c@{}} $n$\\$(\cm-3)$ \end{tabular} & \begin{tabular}{@{}c@{}} $P/k_B$\\($\cm-3$K) \end{tabular} &$f_V$& $f_{m}$ & \begin{tabular}{@{}c@{}} $n$\\$(\cm-3)$ \end{tabular} & \begin{tabular}{@{}c@{}}  $P/k_B$\\($\cm-3$K) \end{tabular} & $f_{m}$  \\ 
  \hline
   ``fvh0.60a'' &  3&  133 & 4.0E3 &  0.006 & 0.833  &12  & 0.4 &  4.0E3 & 0.394& 0.165 & 4.0E-3 &4.0E3 & 0.002 \\ 
   ``fvh0.90a'' &2.5 & 100 & 1.0E4 & 0.011& 0.923    &5 & 1.0  & 1.0E4 & 0.089  & 0.075 & 3.2E-3 &1.0E4 & 0.002 \\
   \hline
  \end{tabular}
\vspace{0.3in}  \\
  (b) Self-consistent models \\
  \begin{tabular}{c |     c    c  c  c  | c      c     c  c  |  c  c  c}
    \hline
      & &Cold & & & &Warm & && &Hot  &\\
  Run  &  \begin{tabular}{@{}c@{}} $n_V$\\$(\cm-3)$ \end{tabular} & \begin{tabular}{@{}c@{}} $P_V/k_B$\\($\cm-3$K) \end{tabular} & $f_{V}$ & $f_{m}$ & \begin{tabular}{@{}c@{}} $n_V$\\$(\cm-3)$ \end{tabular} & \begin{tabular}{@{}c@{}} $P_V/k_B$\\($\cm-3$K) \end{tabular} &$f_V$& $f_{m}$ &\begin{tabular}{@{}c@{}} $n_V$\\$(\cm-3)$ \end{tabular} & \begin{tabular}{@{}c@{}}  $P_V/k_B$\\($\cm-3$K) \end{tabular} & $f_{m}$ \\ 
  \hline
     ``fvh0.66ss'' & 58&  2.9E3 & 0.015 & 0.847  & 0.5  & 3.0E3 &  0.322 & 0.152 & 1.1E-3 & 1.9E3 &  0.001 \\ 
   ``fvh0.82ss'' &108 & 3.9E4 & 0.002 & 0.238  & 4.3 & 4.5E4  & 0.177  & 0.756 &  6.7E-3 & 6.2E4 &  0.006  \\
   \hline
  \end{tabular}
\label{t:ISM_para} 
\end{sidewaystable}

We evolve a SNR in four inhomogeneous ISM set-ups, which have the same mean density $1\cm-3$ but different volume fractions of hot gas $f_{\rm{V,hot}}$, including two idealized medium and two self-consistently generated ISM by multiple SN:

i. ``fvh0.60a" (``fvh0.60'' denotes  $f_{\rm{V,hot}}= 0.6 $, and ``a'' artificial) ; 

ii. ``fvh0.90a";  

iii. ``fvh0.66ss'' (``ss'' denotes self-consistent);

iv. ``fvh0.82ss''.

In the artificial ISM models ``fvh0.60a" and ``fvh0.90a", clouds are identical to each other, and each cloud is spherical with a core of cold medium and a layer of warm phase. All three phases are in pressure equilibrium. The warm and cold phases are thermally stable. The size and mass of the clouds in ``fvh0.60a'' are chosen based on the results of the run $\bar{n}1\_ S200$ from Section \ref{sec:III}; whereas ``fvh0.90a'' represents a more extreme model where 90\% of the volume is hot and 90\% of the mass cold. The locations of clouds are random. Detailed parameters of the three phases, such as the cloud radius $R_{\rm{cl}}$, density $n$, pressure $P/k_B$ and volume/mass fraction $f_V/f_m$ are listed in Table \ref{t:ISM_para} (a). The box size is 300 pc, and the resolution is 0.5 pc. The medium is static initially.

For the self-consistently generated ISM, ``fvh0.66ss" is the output from the run $\bar{n}1\_ S200$ of Section \ref{sec:III} at $t=1.1\times 10^8 \yr$, and ``fvh0.82ss'' from $\bar{n}1\_ S3000$ at $t = 7.3\times 10^6 \yr$. Statistics of the ISM such as the volume-weighted density $n_V$, pressure $P_V/k_B$ and $f_V/f_m$ of each phase are shown in Table \ref{t:ISM_para} (b). To avoid the new SNR propagating outside of the simulation domain, we duplicate eight times the periodic box, assembling the eight identical boxes to make a larger one, which is 286 pc on each side, twice the size of the original one. The resolution 1.12 pc are the same as the original runs. 

We start by injecting $10\msun $ mass and $10^{51} \erg$ energy into a sphere with $M_{\rm{init}} \approx 200 \msun$ ($R_{\rm{init}}\approx 12.5\pc$) at the center of the box. For the artificial medium, the initial energy partition is 30\% kinetic plus 70\% thermal. The injected thermal energy and mass are distributed evenly within the sphere. The initial $E_{\rm{k}}$ of each grid is proportional to $r^2$, where $r$ is the distance of grid from the explosion center. We also tried cases where we inject energy as pure thermal or pure kinetic, and the results converge after the blast wave reaches about 25 pc. For a self-consistently generated ISM, the energy added is purely thermal, and the injected mass and energy are uniformly distributed in the initial sphere.

\subsubsection{SN feedback in a medium with \fvh$=0.6$}

We first take a detailed look at the SNR in the ``fvh0.60a'' model, to understand the physics of the evolution in a multiphase medium, and then compare results of all models in this section. 

\begin{sidewaysfigure*}
\begin{center} 
\includegraphics[width=1.0\textwidth]{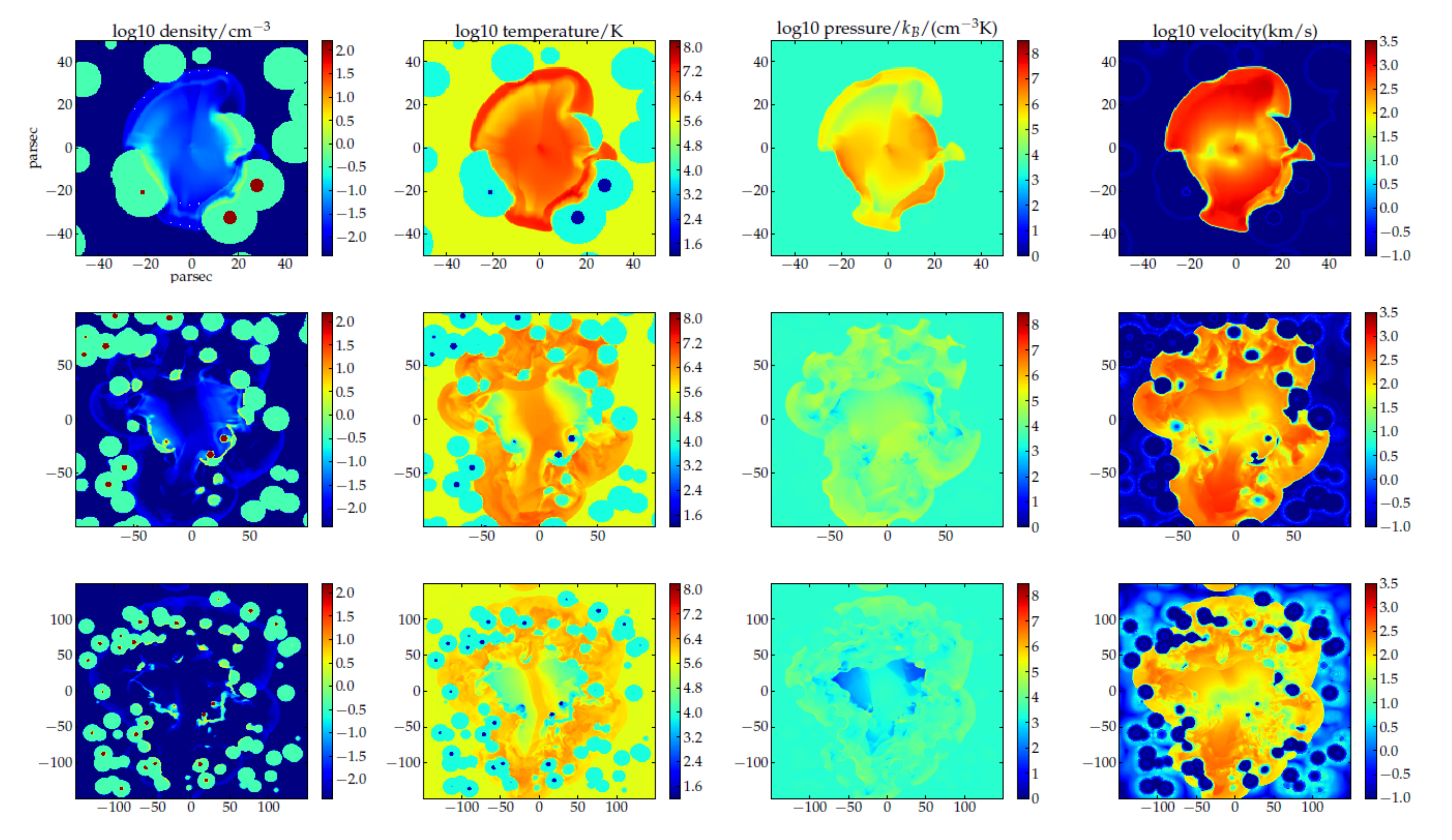}
\caption{Slices of density, temperature, pressure and velocity of a $10^{51}$ erg SNR in ISM model ``fvh0.60a" with $\bar{n}=1 \cm-3$ (see Table \ref{t:ISM_para} for full parameters). Snapshots are taken at $1.9\times 10^4$ year (upper panels), $1.1\times 10^5$ year (middle) and $2.6\times 10^5$ year (lower), respectively. The velocity plot has a floor of 0.1 km/s for log-scale display. Note that the first two rows have zoomed-in views. }
\label{f:2d_fvh06}
\end{center}
\end{sidewaysfigure*}

Fig. \ref{f:2d_fvh06} shows 2D snapshots centered on the SNR in the ``fvh0.60a" model at $t=1.9\times 10^4, 1.1\times 10^5$ and $2.6 \times 10^5$ years, respectively. The first two rows use zoomed-in views. Visually the evolution is very different from that in the uniform ISM model. At $t =  1.9\times 10^4$ year, part of the blast wave has run into warm clouds where the shock velocity slows down. The geometry of the remnant thus deviates from spherical symmetry. The blast wave continues to travel fast in the hot inter-cloud medium -- the least obstructed channel. At the same time, the shock heats and strips the clouds. The warm clouds are relatively easy to disrupt and accelerate, and are mixed with the hot component. The dense cold clouds remain largely intact and static due to their larger inertia and smaller cross sections. There is a visible density enhancement at the shock front in the hot gas, but a dense and cool ($\sim 10^4$ K) shell, which features the ``pressure-driven snowplow'' phase of a SNR in a uniform medium, never forms. The post-shock region is also dissimilar to the Sedov-Taylor solution: the physical quantities inside the blast wave domain (excluding the clouds) are nearly uniform -- it does not show the typical radial velocity distribution as $v \propto R$, nor the progressively lower density and higher temperature toward the center of the blast wave.  The shocked regions are turbulent, as a result of the violent interaction between the blast wave and the embedded clouds. At the end of the simulation, the warm clouds within $R\sim 50\pc $ are largely disrupted, while the cold cores have survived. Clouds in the outer part of the SNR are less distorted, as the power of the blast wave subsides when the energy is spread into a larger volume. 
 
\begin{sidewaysfigure*}
\begin{center}
\includegraphics[width=1.1\textwidth]{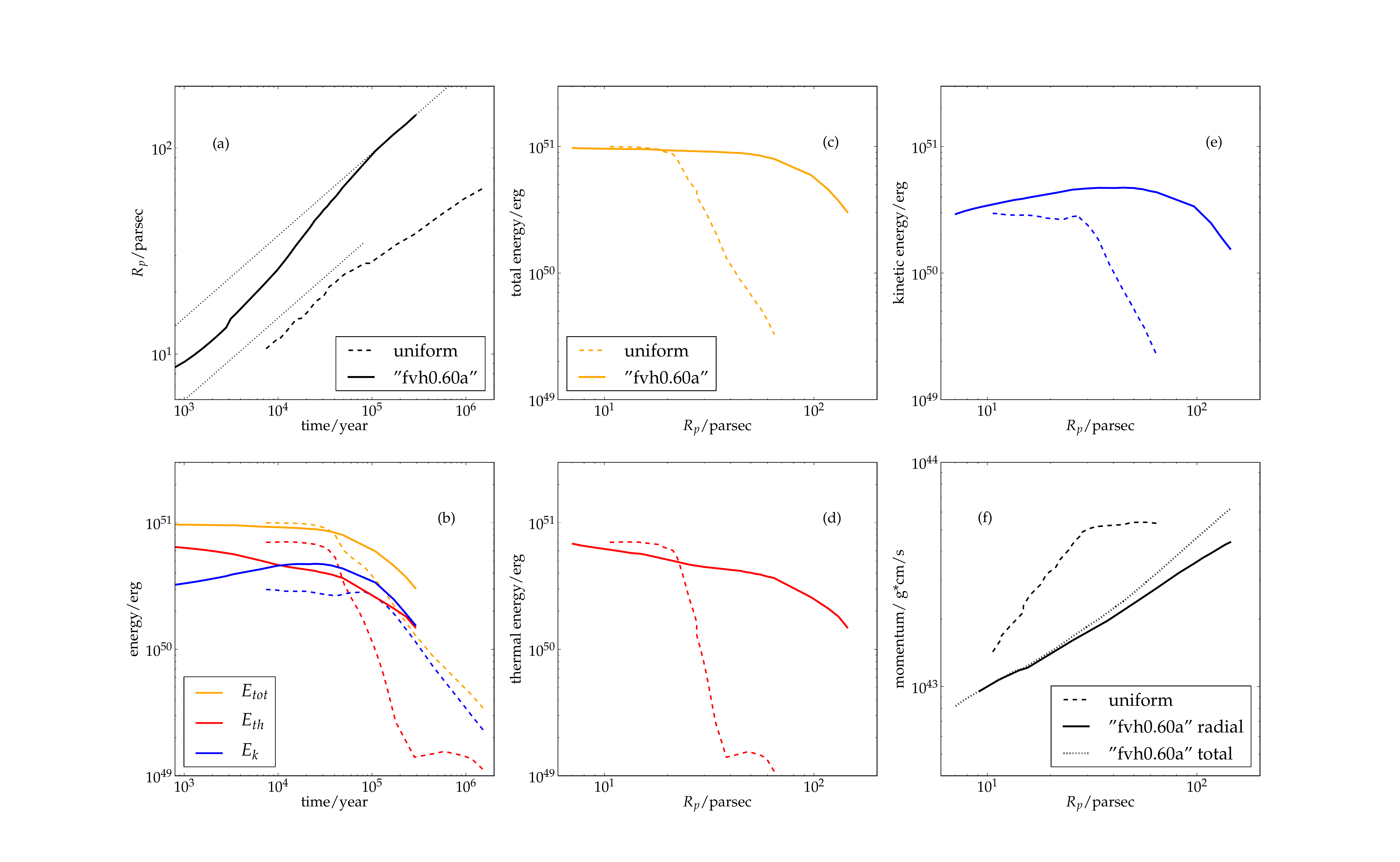}
\caption{Evolution of a $10^{51}$ erg SNR in the ``fvh0.60a" multiphase ISM model with $\bar{n}= 1\cm-3$. (a) radius of the blast wave $R_p$ vs time; (b) energy ($E_{\rm{tot}}, E_{\rm{th}}, E_{\rm{k}}$) vs time; (c)      
$E_{\rm{tot}}$ vs $R_p$; (d) $E_{\rm{th}}$ vs $R_p$; (e) $E_{\rm{k}}$ vs $R_p$; (f) momentum vs $R_p$. Solid lines: ``fvh0.60a" model; dashed lines: the uniform model with the same mean density. The dotted lines in (a) show $R_p -t$ in a uniform medium with $n = 0.4 \cm-3$ and $0.004 \cm-3$, respectively, as predicted by the ST solution up to their separate $t_{\rm{cool}}$ (Eq. \ref{eq:tcool}). In (f), the dotted and solid lines are for the total momentum and its radial component, respectively. }
\label{f:e_r_mom_fvh06}
\end{center}
\end{sidewaysfigure*}

Fig. \ref{f:e_r_mom_fvh06} shows the radial position of the shock $R_p$ vs time and the energy/momentum feedback as a function of time and $R_p$. For comparison, the dashed lines indicate the SNR evolution in a uniform model with the same mean density $n=1\cm-3$, same as what is shown in Fig. \ref{f:e_r_mom}. It is obvious that the evolution in a HDMM is very different from that in a uniform case.

Panel (a) of Fig. \ref{f:e_r_mom_fvh06} shows the evolution of $R_p$ vs $t$. We determine $R_p$ in the following way: since initially the medium is static, we divide the simulation domain into concentric spherical shells centered on the SN explosion site with the thickness of the cell size, and define $R_p$ as the radius beyond which no shells has more than 10\% of cells with $v > 10$ km/s. From the plot, the blast wave travels much faster and further in the multiphase medium. For the uniform case, the power-law index of $R_p$-$t$ turns from $0.4$ to $\sim 0.29$ after the SNR enters the ``pressure-driven snowplow" phase; in contrast, in the multiphase model  is more close to a single power-law function, with a slope $\approx$ 0.55. As a result, by $t=1.8 \times 10^5$ year, the difference in shock radius is more than a factor of 3, and the volume enclosed differs by a factor of $\sim 30$. The hot-dominated medium transfers the blast wave out much more efficiently. 

To see the importance of the clouds in determining the blast wave velocity, we plotted $R_p-t$ for a uniform medium with $n = n_{\rm{hot}} = 4\times 10^{-3} \cm-3$ and $n=n_{\rm{warm}} = 0.4\cm-3$, respectively (dotted lines in panel(a) of Fig. \ref{f:e_r_mom_fvh06}), as predicted by the ST solution up to their separate ${t_{\rm{cool}}}$ (Eq. \ref{eq:tcool}).  At $t \lesssim 10^5$ year, the curve for ``fvh0.60a" lies in between the above cases; after that, the evolution almost coincides with that in a $n = 4\times 10^{-3} \cm-3$ medium without energy loss. This is because that during the early stages, the blast wave is powerful and interacts more with clouds, thus the velocity is reduced, whereas at later stages, the shock largely bypass the clouds and travels more freely in the hot medium. This agrees with what we have seen in the slice snapshots. 

We estimate what fraction of a cloud interacts with the blast wave. For a cloud of the size $2R_{\rm{cl}} $, density $\rho_{\rm{cl}}$ and pressure $P_{\rm{cl}}$ that is located at a distance $d_{\rm{cl}}$ from the center of a spherical blast wave, at a time t, the length scale of the cloud passed by the shock is $R_{\rm{sh}}(t)=\int^{t}_{ t_0(R_p=d_{\rm{cl}}) } v_{\rm{sh}}(t') dt'$, where the shock velocity in the cloud $v_{\rm{sh}}(t) \sim \sqrt{P_{\rm{sh}} (t)/\rho_{\rm{cl}}} \sim c_{\rm{s,cl}} \sqrt{P_{\rm{sh}}(t)/P_{\rm{cl}}}$, $P_{\rm{sh}}$ is downstream pressure of the shock,  and $c_{\rm{s,cl}}$ the sound speed of the pre-shock cloud \citep{klein94}. Assuming $P_{\rm{sh}}$  is uniform within the SNR and $\propto R_p^{-3}$, and $R_p \propto t^{\eta}$, then at $t=t_{\rm{fade}}$, when the blast wave decays to a sound wave, i.e. when $P_{\rm{sh}}\approx P_{\rm{cl}}$, we have
\begin{equation}
\frac{R_{\rm{sh}}}{2R_{\rm{cl}}} = \frac{2}{2-3\eta}\  (\frac{t_{\rm{fade}}}{t_{\rm{s,cl}}})\ [1 - (\frac{d_{\rm{cl}}}{R_{\rm{fade}}} )^{(2-3\eta)/2\eta}],
\label{eq:r_sh}
\end{equation}
where $R_{\rm{sh}}/2R_{\rm{cl}}$ is the fraction of a cloud being shocked, $t_{\rm{s,cl}}$ is the sound crossing time of the cloud ($\equiv 2R_{\rm{cl}}/c_{\rm{s,cl}}$), and $R_{\rm{fade}}$ is the radius of the blast wave at $t_{\rm{fade}}$. Note that the closer a cloud is to the center of the SNR, the larger $R_{\rm{sh}}/2R_{\rm{cl}}$ it has.  Given the approximate value $\eta \sim 0.55$, $R_{\rm{fade}} \sim 200 \pc $, we have $R_{\rm{sh}}/2R_{\rm{cl}} \approx 0.5$ and 1.2 for the cold and warm cloud, respectively, at $d_{\rm{cl}} = 0$. This means that only the warm clouds near the center can fully interact with the blast wave, and cold clouds are mostly bypassed. In other words, for most cold clouds encompassed by a SNR, the interior of the clump does not even know the existence of the blast wave before it fades away. 

Panel (b) of Fig. \ref{f:e_r_mom_fvh06} shows the SN energy as a function of time. The SN energy is obtained by summing up all energy within $R_p$ with the energy from the enclosed ISM deducted. The ISM energy density is acquired by averaging a sample outside of the SNR. The evolution of energy has two stages: an energy conserving phase followed by a radiative phase, divided by when roughly $20\%$ of total energy is lost. For ``fvh0.60a", this transition happens at $t\sim 5 \times 10^4$ year and $R_p \sim 65\pc$. The time is similar to that in the uniform case, while the corresponding $R_p$ is about 2.7 times larger. The energy feedback shows interesting contrasts to the uniform case for both stages. Initially, after the deposition of energy as 30\% kinetic and 70\% thermal, $E_k$ rises while $E_{\rm{th}}$ declines, deviating from the ST solution. $E_{\rm{k}}$ reaches its maximum of about 40\% of SN energy  at approximately the end of the energy conserving phase. The results do not show qualitative variation when we change the initial energy deposition to $100\%$ thermal or $100\%$ kinetic -- $E_{\rm{k}}$ always exceeds that in the ST solution, though $E_k/E_{\rm{tot}}$ when $E_{\rm{k}}$ peaks can vary by about $20\%$. This initial conversion of thermal to kinetic energy is reminiscent of free expansion, and is most likely due to expansion of the shock-heated clouds into the tenuous medium.

Despite that the decline of \eth happens earlier than the uniform case, the rate at which it decreases is much smaller in a HDMM. While in the uniform medium, \eth drops by a factor of 50 within $2\times 10^5$ years, the remnant in ``fvh0.60a" loses only about 50\%. The main reason for the significantly reduced cooling rate is that the blast tends to choose a more tenuous path, which is cooling-inefficient, and largely bypassing the dense clouds. Another reason for the mild loss rate of $E_{\rm{th}}$ is that at later stages $E_{\rm{k}}$ can convert to $E_{\rm{th}}$ through turbulence. $E_{\rm{k}}$, on the other hand, does not show very different time evolution from that in the uniform phase. 

The plots of energy vs $R_p$ (panel (b)-(d)) contain no new information but simply re-present the data from $R_p-t$ and $E-t$. It is obvious that the energy is transported to a much larger volume by the fast-moving blast wave, and drops much more mildly with $R_p$ in a HDMM, especially for $E_{\rm{th}}$. 
In a uniform medium, SN energy is negligible after the catastrophic cooling at $R_p = 65 \pc$, whereas in a multiphase ISM, when the shock fades to a sound wave at $R_p>130\pc$, $\sim 25\%$ of the SN energy is still maintained in the ``fvh0.60a" model.

Momentum $\mathcal{P}$ as a function of $R_p$ is shown in panel (f) of Fig. \ref{f:e_r_mom_fvh06}. The momentum does not increase as fast with $R_p$ in a multiphase ISM. Given $\mathcal{P} = \sqrt{2mE_{\rm{k}}}$, the smaller $\mathcal{P}$ and larger $E_{\rm{k}}$ implies much less mass is involved in sharing the kinetic energy in a HDMM. Note that no thin shell forms at the shock front which carries most of the radial momentum, as happens in a uniform medium. At the end of the simulation, the radial momentum in ``fvh0.60a" is about $4.3\times 10^ {43}$ g$\cdot$cm/s, and is still rising. But it is not likely to keep increasing for long, since the post-shock pressure is about to drop to that of the ISM and will no longer convert into the outgoing momentum. This suggests that the total momentum injected in a clumpy medium is close to that in the uniform case $5.3\times 10^{43}$ g$\cdot$cm/s. We also plot the total momentum (dotted line), which initially coincides with the radial momentum, but later exceeds it since a non-radial component develops as the blast wave interacts with the clouds. At the end of the simulation, the total momentum is roughly 30\% larger than the radial one in ``fvh0.60a".

\subsubsection{SN feedback as a function of \fvh} 
\label{sec:subgrid}

In this subsection, we compare SNR evolution in all 4 multiphase models, and present the fitting formulae to quantify SN feedback in a HDMM.  
The fitting formulae in this section are for $E_{\rm{SN}}=10^{51}$ erg, and gas average density $\bar{n}=1\cm-3$ and \fvh$\gtrsim 0.5$. 

\begin{sidewaysfigure*}
\begin{center} 
\includegraphics[width=1.0\textwidth]{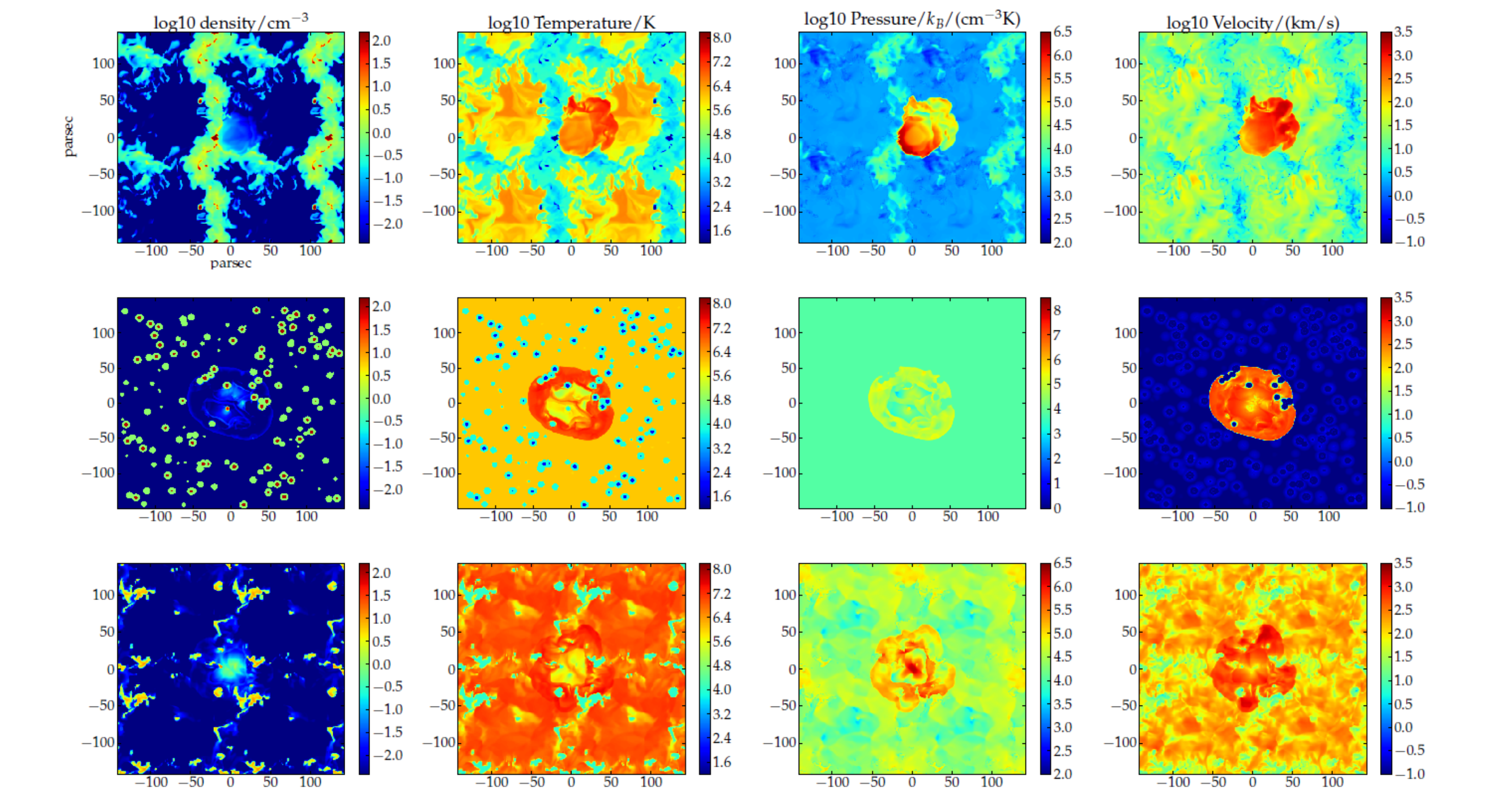}
\caption{ 
Snapshots for a $10^{51}$ erg SNR evolving in ISM models ``fvh0.66ss'' (upper row), ``fvh0.90a'' (middle), ``fvh0.82ss''(lower), respectively, at $3 \times 10^4$ year. See Section \ref{sec:IIpara} for model details. The velocity plot has a floor of 0.1 km/s for log-scale display.  
}
\label{f:2d_all_multi}
\end{center}
\end{sidewaysfigure*}

Fig. \ref{f:2d_all_multi} shows the snapshots of the three multiphase ISM models  ``fvh0.66ss'', ``fvh0.90a'' and ``fvh0.82ss'' at $t = 3\times 10^4$ year. It is clear that the blast wave travels mostly in the hot phase. For ``fvh0.66ss'' model, where the clouds form walls and hot bubbles do not fully connect to each other, the shock goes further in the low-density bubble and deviates from spherical symmetry. Within the SNR, the pressure is higher toward the clouds. For ``fvh0.90a'' and ``fvh0.82ss'', where the clouds are smaller with nearly random locations, the blast waves are close to spherical.

\begin{figure}
\begin{center}
\includegraphics[width=0.8\textwidth]{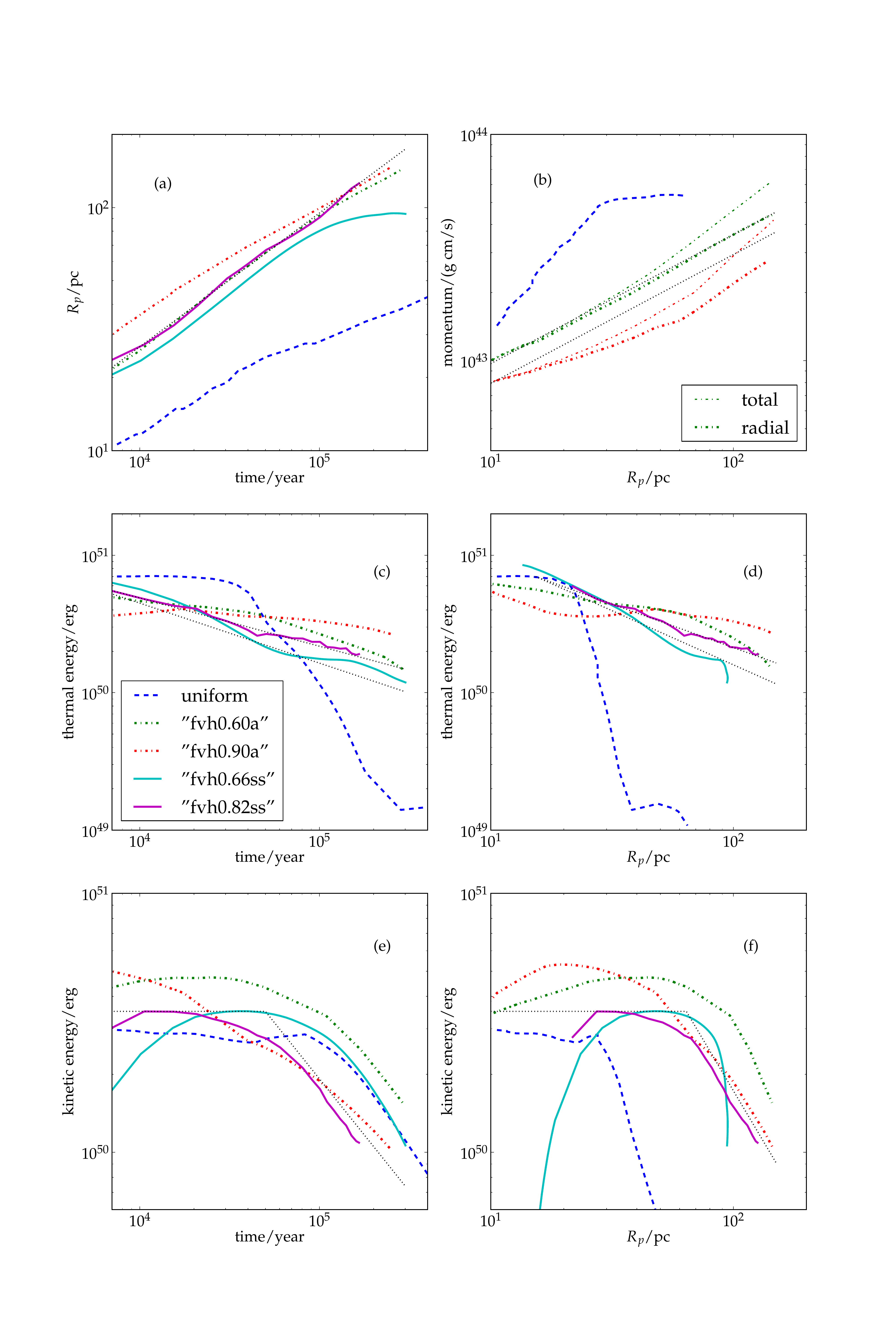}
\caption{SN feedback in the 4 multiphase ISM models described in Section \ref{sec:II} vs a uniform medium with the same mean density $n=1\cm-3$. Line styles and the corresponding models are given in the legend in panel (c). (a) $R_p$ vs time; (b) momentum vs $R_p$; (c) $E_{\rm{th}}$ vs time; (d) $E_{\rm{th}}$ vs $R_p$; (e) $E_{\rm{k}}$ vs time; (f) $E_{\rm{k}}$ vs $R_p$. In (b), only results of ``fvh0.60a'' and ``fvh0.82a'' are shown, and the thin and thick lines denote the total momentum and the radial component, respectively. The black dotted lines in each panel indicate the fitting formula presented in Section \ref{sec:subgrid}.  }
\label{f:E_t_allmulti}
\end{center}
\end{figure}

Fig. \ref{f:E_t_allmulti} compares SN feedback in all multiphase ISM models. For reference, the uniform case is shown with the blue dashed lines. For ``fvh0.90a'', $R_p$, \eth, \ek are determined the same way as for ``fvh0.60a''. For the self-consistently generated ISM models, which are turbulent and the SNR may have an irregular shape, we briefly describe how we determine the size and energy of the SNR. For ``fvh0.66ss'', we adopt the following criterion to determine if a cell is within the SNR. A post-shock cell should satisfy at least one of the two conditions: a. $v>150$ km/s; b. $nT>10^{4.5}\cm-3$ K. The adopted velocity and pressure floors are beyond (but very close to) the maximum values in the pre-SN ISM. We then define the ``effective radius'' $R_p \equiv (3V_{\rm{shock}}/4\pi)^{1/3}$, where $V_{\rm{shock}}$ is the total volume of the post-shock region. For the ``fvh0.82ss'' model, since the ISM has higher pressure and velocities, and the spatial variations are larger than ``fvh0.66ss'', it is difficult to obtain $R_p$  in the same way. Taking advantage of the relatively regular shape of the SNR, we take snapshots and identify the $R_p$ by eye. The SN energy is determined by running a reference simulation for each self-consistent ISM model where no SN is added. At each time step, the difference of $E_k$ and $E_{\rm{th}}$ between the two simulations are deemed those associated with the SN. 

In Fig. \ref{f:E_t_allmulti}, one may first notice that the four multiphase models show roughly similar SNR evolution, but they are quite different from the uniform model (except panel (e) $E_k$-t). The comparable evolution is interesting because the four multiphase models have fairly different configurations: cloud shape, $f_{\rm{V,hot}}$, turbulence, and even resolution. This implies that those differences are secondary to the fact that they share the same feature of a HDMM, distinct from the uniform case. 

Let us then look at the feedback in some details. Panel (a) of Fig. \ref{f:E_t_allmulti} shows $R_p$ vs $t$. SNRs in all multiphase ISM models go much faster and further than the uniform medium. For both the artificial and self-consistent ISM, $R_p$ is larger for higher \fvh at a given time, but overall the difference is small. For ``fvh0.66ss'', at later stages, $R_p$ reaches its maximum of 95 pc, when the shock subsides and merges with the cloud boundaries of the bubbles. The SN feedback is confined by the walls, thus it does not influence as large volume as when hot gas has a connecting topology. Overall, $R_p$ vs $t$ is roughly represented by the following power-law function
\begin{equation}
R_p = 22\pc (\frac{t}{7\times 10^3 \yr})^{0.55} .
\label{eq:r-t_m}
\end{equation}
The power-law index, 0.55, is larger than the ST solution. The corresponding curve is shown with the black dotted line (same linestyles are adopted for other fitting formulae presented in this subsection in Fig. \ref{f:E_t_allmulti}). 

Panel (c) of Fig. \ref{f:E_t_allmulti} shows the time evolution of $E_{\rm{th}}$, which is characterized by an early onset of decline and a small cooling rate for all multiphase models. Self-consistent ISM models lose \eth at a somewhat larger rate,  which may be due to their turbulent nature that enhances the dissipation rate. If we compare the two artificial ISM models or the self-consistent ones, a larger \fvh yields a smaller cooling rate. $E_{\rm{th}}(t)$ has a power-law-like form, with power-law indices (absolute value) much smaller than the uniform case. We adopt a very simple formula to fit the power-law indices for the two self-consistent models, which are inversely proportional to $f_{\rm{V,hot}}$:
\begin{equation}
E_{\rm{th}} (t) = \begin{cases}
0.65E_{\rm{SN}}& \text{for $t\leq 0.09t_{\rm{cool,\bar{n}}}$},\\
0.65E_{\rm{SN}}(\dfrac{t}{0.09t_{\rm{cool,\bar{n}}}})^{-0.36\times (0.8/f_{\rm{V,hot}})}\  & \text{for $t>0.09t_{\rm{cool,\bar{n}}} $ },
\end{cases}
\label{eq:eth-t_m}
\end{equation}
where $t_{\rm{cool,\bar{n}}}=4.9 \times 10^4\yr (\bar{n}/1\cm-3)^{-0.55} (E_{\rm{SN}}/10^{51}\erg)^{0.22}$, following Eq. \ref{eq:tcool}. 
Note that the functional form of the power-law index is chosen somewhat arbitrarily, with no direct physcial motivation. Only two values of $f_{\rm{V,hot}}$ are fit, meaning that it is not strongly constrained; however, this is probably only of secondary importance for sub-grid modeling, as $f_{\rm{V,hot}}$ does not vary strongly across HDMM conditions. We set a ceiling for it at 65\% of the SN energy, to avoid the divergence at $t=0$. The onset of cooling, which is about 10\% of $t_{\rm{cool,\bar{n}}}$, is much earlier than the uniform case, while the canonical value of the power-law index of $E_{\rm{th}} -t$ for a HDMM, -0.36, is significantly smaller than the uniform case, $\approx -2$.
 
Panel (d) of Fig. \ref{f:E_t_allmulti} shows \eth vs $R_p$.
The fitting formula can be obtained from combining Eq. \ref{eq:r-t_m} and \ref{eq:eth-t_m}:
\begin{equation}
E_{\rm{th}} (R_p) = \begin{cases}
0.65E_{\rm{SN}}& \text{for $R_p\leq 0.73 R_{\rm{cool,\bar{n}}}$,}\\
0.65E_{\rm{SN}}(\dfrac{R_p}{0.73 R_{\rm{cool,\bar{n}}}})^{-0.65\times (0.8/f_{\rm{V,hot}})} & \text{for $R_p>0.73 R_{\rm{cool,\bar{n}}}$},
\end{cases}
\end{equation}
where $R_{\rm{cool,\bar{n}}}=23.7\pc (\bar{n}/1\cm-3)^{-0.42} (E_{\rm{SN}}/10^{51}\erg)^{0.29}$, following Eq. \ref{eq:rcool}. Again, the SN energy starts to cool at a somewhat smaller $R_p$, while the dependence of $E_{\rm{th}}$ on $R_p$ is much less sensitive in a HDMM than in a uniform medium, which has a power-law index of about -7. Note that the evolution of \eth in our three-phase ISM simulations is different from that in two-phase models. \cite{martizzi15} find that the the power-law slope of the declining phase of \eth vs $R_p$ is similar to the uniform case (see their Fig 4). KO15 show that the energy evolution in a two-phase ISM is relatively close to the uniform case with the warm phase density (see their Fig 10), since the cold clumps only have a small volume fraction. In a three-phase medium, we find that evolution is more complicated. Although the late evolution of $R_p$ is close to that in the hot medium only, the energy evolution cannot be rescaled to the uniform case with any density. It thus suggests that some new model is needed which takes into account both hot gas and warm clouds.

Panel (e) of Fig. \ref{f:E_t_allmulti} shows the time evolution of \ek, which in multiphase media does not show large deviations from that in the uniform medium, although in the energy conserving phase,  \ek has a higher fraction of SN energy than 30\%. 
A simple fitting formula is 
\begin{equation}
E_k(t) = \begin{cases}
0.35E_{\rm{SN}} & \text{for $t\leq t_{\rm{cool,\bar{n}}}$},\\
0.35E_{\rm{SN}}(\dfrac{t}{t_{\rm{cool,\bar{n}}}})^{-0.87} & \text{for $t>t_{\rm{cool,\bar{n}}}$}.
\end{cases}
\label{eq:ek-t_m}
\end{equation}

$E_k$ vs $R_p$ is shown in panel (f) of Fig \ref{f:E_t_allmulti}. Similar to \eth, \ek is conveyed to a much larger volume in a HDMM. For the fitting formula, combining Eq. \ref{eq:r-t_m} and \ref{eq:ek-t_m}, we get
\begin{equation}
E_k (R_p) = \begin{cases}
0.35E_{\rm{SN}} & \text{for $R_p\leq 2.7R_{\rm{cool,\bar{n}}}$},\\
0.35E_{\rm{SN}}(\dfrac{R_p}{2.7R_{\rm{cool,\bar{n}}}})^{-1.58}  & \text{for $R_p>2.7R_{\rm{cool,\bar{n}}}$}.
\end{cases}
\label{eq:ek-R_m}
\end{equation} 
Note that in the declining phase, the power-law index of $R_p$, -1.58, is less steep than the uniform case, -3, consistent with that the momentum is not conserved, but still rising when $E_k$ declines.

The momentum evolution is shown in panel (b) of Fig \ref{f:E_t_allmulti}. Both the total momentum and the radial component are presented. We only plot the artificial ISM cases, since in a turbulent medium it is hard to isolate the momentum directly related to the new SNR. The SNR in the ``fvh0.90a'' model has a consistently smaller momentum than that in ``fvh0.60a'', at a given an $R_p$, since less mass is involved in the blast wave in a more hot-dominated medium. As the shock goes to further distances,  momentum is developed in non-radial directions and the post-shock region becomes more turbulent. For the most part of $R_p$, the momentum in a multiphase medium is much smaller than the uniform case, but the terminal momenta are likely to be similar. For the radial momentum, we adopt the following fitting formula: a power-law increase with $R_p$ followed by a plateau as in the uniform case:
\begin{equation}
\mathcal{P}_{\rm{rad}} = \begin{cases}
5\times 10^{43} \rm{g\cdot cm\ s^{-1}} (\frac{R_p}{7.5R_{\rm{cool,\bar{n}} }})^{0.57} (\frac{f_{\rm{V,hot}}}{0.6})^{-0.5} & \text{for $R_p \leq 7.5R_{\rm{cool,\bar{n}}}(\frac{f_{\rm{V,hot}}}{0.6})^{0.88} $},\\
 5\times 10^{43} \rm{g\cdot cm\ s^{-1}} & \text{for $R_p > 7.5R_{\rm{cool,\bar{n}}}(\frac{f_{\rm{V,hot}}}{0.6})^{0.88} $}. 
\end{cases}
\end{equation}
The power-law indices of $E_{\rm{SN}}$ and $\bar{n}$ follow those of the uniform case (Eq. \ref{eq:P_end}). Note that while a single power-law describes the evolution in ``fvh0.60a'' almost perfectly, it does not agree well with ``fvh0.90a'', which may be better fitted by a broken power-law. Here we simply make \fvh a factor that only influences the zero point of curve, to reflect that less momentum is developed at a given $R_p$ in a medium with a larger \fvh.

Note that we have shown the uniform case with the same mean density of the HDMM models. Since the cold phase that contains most mass actually fills little of the volume, from a physical perspective, one could also compare to uniform models with the same \textit{median} density. We do not show them here to avoid overcrowding the figures. However, as we have shown in Section \ref{sec:I}, the SN feedback in a uniform medium is well described by power-law functions, and the power-law indices vary little for different gas densities. So as a rule of thumb, the feedback curves simply shift horizontally for different densities.

\vspace{0.2 in}

In summary, the evolution of a SNR in a HDMM is characterized by
a much larger impact volume, a reduced cooling rate, and a slowly developed momentum. When the medium is uniform in density, there is no preferential path and the shock encounters mass at the same rate in all directions; when the medium is clumpy, however, the shock tends to choose the more tenuous regions where it travels much faster and further. The connected rarefied channels give vent to the blast wave energy, which largely goes around the clouds at later stages. Consequently, not all mass within $R_p$ interacts significantly with the blast wave. A HDMM with connecting tenuous tunnels therefore greatly facilitates SN feedback. In cosmological simulations which cannot resolve the multiphase, turbulent medium, it has been commonly assumed that SN explode in a uniform medium with the mean cell density. The cooling rate can thus be severely overestimated. A more realistic and self-consistent model should consider a SN-modified multiphase ISM.

We can define the domain of influence of a SN in 4D spacetime: 
\begin{equation}
V_{\rm{4D}} =  \frac{4\pi}{3} R_{\rm{0.2}}^3 t_{\rm{0.2}}.
\end{equation}
If we define $ R_{\rm{0.2}}$ and $ t_{\rm{0.2}}$ as the $R_p$ and time when 20\% of SN is energy is left, then for a $10^{51}$ erg SN, 
$R_{\rm{0.2}}=33.8 \pc$ and $t_{\rm{0.2}}=1.7\times 10^5 \yr$ in a uniform medium with $1\cm-3$, and $R_{\rm{0.2}}= 190.0\pc$ and $t_{\rm{0.2}}= 3.5\times 10^5\yr$ in a multiphase medium with same mean density and \fvh$=80\%$, according to Eq. \ref{eq:eth-t_m} - \ref{eq:ek-R_m}. The impact volume is therefore 178 times larger and time 2.1 times longer, making a 374 times bigger $V_{\rm{4D}}$ in a HDMM than in a uniform medium.

In real galaxies, it is usually the collective effect of many SN that shape the ISM and launch galactic winds. A larger $V_{\rm{4D}}$ for a SNR means that a new SN is more likely to explode within the $V_{\rm{4D}}$ of an old one.  Thus the chance for SNRs to overlap significantly increases, and energy from multiple SN are ``grouped", which may result in a big runaway explosion. It is therefore very interesting to investigate the question ``Under what circumstances will a HDMM form?", which we will do in the next section.

\section{Multiple SN shaping multiphase ISM} 
\label{sec:III}

In this section, we allow multiple SN to shape the ISM. We experiment with two parameters, the mean density $\bar{n}$ and the SN rate $S$, and analyze the resultant ISM. In particular, we want to find out when there will be a HDMM, which has the potential to drive a wind. 

\subsection{Model Description} 
\label{IIImodel}

The gas is initially uniform with a density $\bar{n}$, temperature $3\times 10^3 $ K and zero velocity. Periodic boundary conditions are applied. SN are assumed to have random locations (we test how good this assumption is later in Section \ref{sec:random}, taking into account runaway OB stars). We explore the parameter space of $(\bar{n},S)$ to study the resultant ISM systematically. Each simulation box contains $128^3$ cells. The box size of each simulation varies with $\bar{n}$: the length is chosen to be $6 R_{\rm{cool}}$ (Eq. \ref{eq:rcool}). The adaptive box size allows us to achieve higher resolution for a denser ISM, thus better capturing the evolution of SNRs and avoiding overcooling problems, which are frequently encountered in coarse resolution simulations. 
 
Each supernova is injected as $10^{51} \erg$ thermal energy and $10 \msun$ mass, which are distributed evenly within a sphere. We choose the initial radius such that the enclosed mass is close to but less than $M_{\rm{init}}$, and we adopt $M_{\rm{init}} = 200 \msun$ for $\bar{n}\le 1\cm-3$ and $100\msun$ for $\bar{n}> 1\cm-3$. As for the duration of the simulation, we note that there is an intrinsic timescale after which ISM has been fully bombarded by random-located SNR, so that the initial uniform distribution of gas is destroyed entirely. This is roughly \begin{equation}
t_{b}=\frac{1}{S\ (2R_{\rm{cool}})^3} = 9.2\ \rm{Myr}\ (\frac{S}{1000\kpc^{-3}\myr-1})^{-1}(\frac{\bar{n}}{1\cm-3})^{1.26}.
\label{eq:tb}
\end{equation} 
We therefore choose the simulation time to be $t_{\rm{sim}}=2.5\ t_b$, so at the end of each simulation $\sim$ 66 SN have exploded in the box.
The time span corresponds to about 10\% of the gas depletion time ($\equiv \bar{n} m_{\rm{H}}/(S\cdot 100\msun)$) for $\bar{n}=1\cm-3$  runs, and 40\% for $\bar{n}=30\cm-3$. We find that the ISM reaches steady state (if there is one) long before the end of the simulation.
 
We assume PEH rate scale linearly with the $S$, and adopt the following formula for a uniform PEH rate per H atom \citep{draine11}:
\begin{equation}
\Gamma_{\rm{pe}} = 1.4\times 10^{-26} \erg\cdot \rm{s^{-1}} ( \frac{S}{100 \kpc^{-3} \myr-1}).
\end{equation}
We implicitly assume a constant dust/gas ratio, in line with a constant metallicity. Given a nominal dust photoabsorption cross section of FUV photons per H atom, $10^{-21} \rm{cm^{2}}$ \citep{draine11}, we find that most of the simulation boxes are in the optically thin regime, except for the $\bar{n} =30\cm-3$ runs where the total optical depth across the box is about 3.3. Therefore a linear scaling relation of PEH rate with S is, in general, not a bad approximation. 

\subsection{Choice of Parameters}
\label{sec:para}

The $(\bar{n},S)$ combinations encountered in different galaxy environments are not well-constrained, and are likely to be time-variable. Although on scales of kpc and above, we have the Kennicutt relation on the star formation (SF) rate and mean gas surface density \citep{kennicutt98} (which has intrinsic scatters itself), on a smaller scale, like that of our simulation boxes, recent observations have shown large deviations from it. For example, by doing stellar counts for the star forming region in the MW, \cite{heiderman10} found that the SF rate can be above the Kennicutt relation by at least an order of magnitude for $\Sigma_{\rm{gas}} \sim 100 \msun\pc^{-2}$ (although this is for small individual star-forming regions, not for large scale ISM). Besides, the Kennicutt relation uses the observables -- the column densities of gas and SF rate, instead of the volume densities, and the scale height of the gas for external galaxies are usually not well-known. Furthermore, Type Ia SN, which are associated with the old stellar populations, do not correlate with gas properties. Lastly, Type II SN, because of the non-negligible velocities of their progenitor OB stars, can migrate out of the SF sites. All these contribute to the scatter of $\bar{n}-S$ relation in real galaxies on different length scales.

Given our incomplete knowledge and the potentially big scatter in the relation of $\bar{n}$ and $S$, we treat the two as free parameters, the relation between them being determined by physical modeling that depends on self-gravity, which has been ignored in this work (see, for example, \cite{kim11,kim13} for self-consistent determination of $\Sigma_{\rm{SFR}}$ as a function of $\Sigma_{\rm{gas}}$ and $\Sigma_{\rm{star}}$ for disk galaxies; for elliptical galaxies, SN are mostly Type Ia from old stellar populations, therefore such self-regulations do not exist.). In terms of choosing what parameter space to explore, we try to combine both observationally- and theoretically-motivated regimes. From the observational side, we cover the following regions and their neighbourhood -- (1) MW disk-average ($\bar{n} \approx 1 \cm-3$ and $S \approx 200 \kpc^{-3} \myr-1 $); (2) MW halo: based on vertical distribution of SN and halo gas; (3) Scaling the MW disk-average to ``starburst" regime using the Schmidt relation $S\propto \rho/t_{\rm{dyn}} \propto \bar{n}^{1.5}$. Beyond the above domains, we also expand the parameter space after we analyze the results of the existing runs, to include those that are theoretically interesting. For example, given the density of the ISM, there is a critical SN rate $S_{\rm{crit}}$ above which no steady state exists. We list in Table \ref{t:simIII} the basic parameters for all simulations in this section. The symbolic name of a simulation, for example, $\bar{n}0.1\_ S10$, means $\bar{n} = 0.1 \cm-3$ and $S = 10 \kpc^{-3} \myr-1$. ``Res'' is short for ``Resolution''. 

For the rest of this subsection, we explain how we estimate the $\bar{n}$ - $S$ correlation in the vertical direction of the MW disk. 
Observation of the 21 cm line indicates that the neutral medium lies within a flat layer approximated by:
\begin{equation}
n_{\rm{HI}} = 0.57\cm-3\lbrace0.7\ \rm{exp}[-(z/127\pc)^2] 
+0.19\ \rm{exp}[-(z/318\pc)^2]+
0.11\ \rm{exp} (-|z|/413\pc)\rbrace ,
\label{eq:nH}
\end{equation}
where $z$ is the vertical distance from mid-plane \citep{dickey90}. \cite{gaensler08} deduced from pulsar dispersion measure and diffuse H$\alpha$ emission that the ionized warm medium of the MW follows an exponential function:
\begin{equation}
n_e = 0.014\cm-3 \rm{exp}(-|z|/1830\pc).
\label{eq:ne}
\end{equation}

The hot ionized gas is less constrained, and the density is very low, so we leave it out for this study. The sum of the above phases then yield the vertical distribution of the MW gas. 

The SN frequency for the MW is about 1/60 $\yr ^{-1}$ for core collapse SN and 1/250 year$^{-1}$ for Type Ia \citep{cappellaro97}. Type Ia SN come from the old stellar disk which has a scale height of $\sim 325$ pc, therefore we adopt the following vertical distribution of Type Ia SN, as averaged over the MW disk (assuming 10 kpc radii in size):
\begin{equation}
S_{\rm{Ia}} = 19.6 \kpc^{-3} \myr-1 \rm{exp} (-|z|/325\pc).
\label{eq:Ia}
\end{equation}
For core collapse SN, the vertical distribution is inferred from that of pulsars at birth, which is a sum of two Gaussian functions \citep{narayan90}: 
\begin{equation}
S_{\rm{CC}} = 167 \kpc^{-3}\myr-1 \{ \ 0.75\ \rm{exp}(z/120\pc)^2 + 0.25\ \rm{exp} (z/360\pc)^2\ \}.
\label{eq:cc}
\end{equation}
Putting Eq. \ref{eq:nH}-\ref{eq:cc} together, we get an $\bar{n}-S$ relation along the vertical direction of the MW. This is shown by the blue solid line in Fig. \ref{f:fvh} and \ref{f:Pv} that we will describe later.

\begin{sidewaystable*}  
\small
\begin{center}
\caption{Input/Output parameters of simulations in Section \ref{sec:III})}
  \begin{tabular}{ c| c c c| c |  c |   c |   c   c  c   c  c}
   \hline
   Run   & \begin{tabular}{@{}c@{}} Box size\\(pc) \end{tabular} &  \begin{tabular}{@{}c@{}} Res\\(pc) \end{tabular}  &\begin{tabular}{@{}c@{}} $t_{\rm{sim}}$\\(Myr) \end{tabular} & \begin{tabular}{@{}c@{}} Cold ($<10^3$K)\\ $f_m$/$f_V$/$\mathcal{M}_V$ \end{tabular} &\begin{tabular}{@{}c@{}} Warm \\ $f_m$/$f_V$/$\mathcal{M}_V$ \end{tabular} & \begin{tabular}{@{}c@{}} Hot ($>2\times 10^5$K)\\ $f_m$/$f_V$/$\mathcal{M}_V$ \end{tabular} & \begin{tabular}{@{}c@{}}$P_V/k_B$\\ ($\cm-3$K) \end{tabular} & $\frac{P_{\rm{pred}}}{P_{V}}$ & $f_{\rm{Ek}}$ & \begin{tabular}{@{}c@{}} $t_{\rm{heat}}$\\(Myr) \end{tabular} & \begin{tabular}{@{}c@{}} TR\footnotemark[1] \end{tabular} \\ \hline

$\bar{n}30\_ S2E4$& 35.7  & 0.28  & 72.5 & 0.14 /0.006 /2.16 & 0.86 /0.99 /0.55 & 2E-06 /0.01 /0.10 & 2.6E+05 & 1.2 & 0.28 &0.13 & N \\
$\bar{n}30\_ S4E4$& ..  & ..  & 35.9 & 0.02 /5E-04 /2.35 & 0.98 /0.97 /0.73 & 1E-05 /0.03 /0.22 & 3.2E+05 & 1.0 & 0.36  &0.09 & N\\
$\bar{n}30\_ S8E4$& ..  & ..  & 17.2 & -- /-- /-- & 1.0 /0.89 /1.00 & 1E-04 /0.11 /0.60 & 3.4E+05 & 1.0 & 0.37 &0.07 & N\\
$\bar{n}30\_ S2E5$& ..  & ..  & 7.2 & -- /-- /-- & 1.0 /0.70 /1.35 & 7E-04 /0.30 /0.83 & 4.8E+05 & 0.9 & 0.39 &0.04 & N\\
$\bar{n}30\_ S8E5$& ..  & ..  & 1.8 & -- /-- /-- & 1.0 /0.31 /1.73 & 0.004 /0.69 /0.85 & 1.8E+06 & 0.6 & 0.36 &0.03 & m\\\hline
$\bar{n}10\_ S3000$& 54.3 & 0.42  & 134.8 & 0.57 /0.04 /1.58 & 0.43 /0.93 /0.72 & 1E-05 /0.03 /0.11 & 3.7E+04 & 2.8 & 0.46 &0.14 & N\\
$\bar{n}10\_ S1E4$& ..  & ..  & 41.6 & 0.11 /0.003 /1.92 & 0.89 /0.77 /0.76 & 7E-04 /0.23 /0.67 & 1.1E+05 & 1.2 & 0.37  &0.12& N\\
$\bar{n}10\_ S2E4$& ..  & ..  & 20.6 & 0.02 /3E-04 /2.59 & 0.98 /0.74 /1.01 & 5E-04 /0.26 /0.66 & 1.2E+05 & 1.1 & 0.31  &0.08 & N\\
$\bar{n}10\_ S5E4$& ..  & ..  & 8.3 & -- / -- /-- & 1.0 /0.48 /1.28 & 0.002 /0.52 /0.70 & 2.5E+05 & 0.8 & 0.34  &0.06 & m\\
$\bar{n}10\_ S1E5$& ..  & ..  & 4.1 & -- / --  / -- & 1.0 /0.43 /1.80 & 0.003 /0.57 /0.90 & 3.6E+05 & 0.6 & 0.46 &0.05 & m\\\hline
$\bar{n}3\_ S300$& 90.0 & 0.70 & 301.9 & 0.84 /0.07 /1.92 & 0.16 /0.88 /1.28 & 2E-05 /0.05 /0.12 & 3.2E+03 & 3.7 & 0.60 & 0.20  & N\\
$\bar{n}3\_ S700$& ..  & ..  & 129.4 & 0.67 /0.04 /2.09 & 0.33 /0.79 /1.20 & 2E-04 /0.16 /0.29 & 7.7E+03 & 3.6 & 0.41  &0.21& N\\
$\bar{n}3\_ S1000$& ..  & ..  & 90.6 & 0.63 /0.03 /1.97 & 0.37 /0.67 /1.18 & 4E-04 /0.30 /0.39 & 1.0E+04 & 3.9 & 0.32  &0.19 & N\\
$\bar{n}3\_ S3000$& ..  & ..  & 30.2 & 0.22 /0.006 /1.94 & 0.78 /0.58 /1.14 & 0.001 /0.42 /0.68 & 3.3E+04 & 1.6 & 0.40  &0.14& N\\
$\bar{n}3\_ S5000$& ..  & ..  & 18.1 & 0.04 /9E-04 /2.58 & 0.96 /0.59 /1.16 & 0.001 /0.41 /0.66 & 4.8E+04 & 1.0 & 0.39  &0.12 & N\\
$\bar{n}3\_ S1E4$& ..  & ..  & 9.1 & 7E-03 /6E-05 /2.33 & 0.99 /0.42 /1.11 & 0.003 /0.58 /0.72 & 9.0E+04 & 0.8 & 0.34  &0.10 & m\\\hline
$\bar{n}1\_ S50$& 142.8 & 1.11  & 453.0 & 0.84 /0.08 /2.44 & 0.16 /0.86 /1.40 & 4E-05 /0.07 /0.15 & 9.6E+02 & 2.1 & 0.56  &0.39 & N\\
$\bar{n}1\_ S100$& ..  & ..  & 226.5 & 0.80 /0.05 /2.46 & 0.20 /0.74 /1.44 & 3E-04 /0.21 /0.32 & 1.6E+03 & 2.5 & 0.49 & 0.33  & N\\
$\bar{n}1\_ S200$& ..  & ..  & 107.5 & 0.81 /0.02 /2.27 & 0.19 /0.4 /1.25 & 0.001 /0.58 /0.42 & 4.1E+03 & 1.9 & 0.14  &0.27 & m\\
$\bar{n}1\_ S500$& ..  & ..  & 45.3 & 0.81 /0.008 /2.42 & 0.19 /0.15 /1.03 & 0.002 /0.85 /0.32 & 1.6E+04 & 1.3 & 0.13  &0.28 & Y\\
$\bar{n}1\_ S1000$& ..  & ..  & 22.9 & 0.69 /0.007 /2.11 & 0.30 /0.17 /1.24 & 0.004 /0.82 /0.48 & 2.6E+04 & 1.6 & 0.26 &0.23 & Y\\
$\bar{n}1\_ S3000$& ..  & ..  & 7.5 & 0.23 /0.002 /2.64 & 0.77 /0.18 /1.33 & 0.006 /0.82 /0.51 & 6.5E+04 & 0.8 & 0.23  &0.17 & Y\\\hline
$\bar{n}0.3\_ S10$&  242.8 &  1.90  & 461.0 & 0.80 /0.04 /2.82 & 0.20 /0.76 /1.30 & 3E-04 /0.20 /0.27 & 4.6E+02 & 0.9 & 0.25  &0.76 & N\\
$\bar{n}0.3\_ S40$& ..  & ..  & 115.2 & 0.82 /0.008 /2.79 & 0.18 /0.29 /1.25 & 0.003 /0.70 /0.47 & 2.3E+03 & 0.7 & 0.16  &0.54  & m\\
$\bar{n}0.3\_ S100$& ..  & ..  & 46.1 & 0.86 /0.005 /3.56 & 0.14 /0.08 /0.89 & 0.004 /0.92 /0.31 & 8.3E+03 & 0.5 & 0.09 & 0.55  & Y\\
$\bar{n}0.3\_ S500$& ..  & ..  & 9.2 & 0.25 /0.003 /3.13 & 0.74 /0.16 /1.21 & 0.008 /0.84 /0.47 & 2.1E+04 & 0.9 & 0.18  &0.36   & Y\\\hline
$\bar{n}0.1\_ S10$& 371.4 &  2.90  & 128.9 & 0.69 /0.008 /2.7 & 0.30 /0.38 /0.92 & 0.003 /0.61 /0.46 & 9.1E+02 & 0.4 & 0.18   &0.64 & m\\
$\bar{n}0.1\_ S50$& ..  & ..  & 25.8 & 0.47 /0.006 /3.34 & 0.52 /0.17 /1.21 & 0.009 /0.82 /0.45 & 4.6E+03 & 0.4 & 0.15  &0.61 & Y\\
$\bar{n}0.1\_ S100$& ..  & ..  & 12.9 & 0.13 /0.002 /3.71 & 0.86 /0.20 /1.33 & 0.01 /0.80 /0.51 & 6.7E+03 & 0.6 & 0.20  &0.50 & Y\\
$\bar{n}0.1\_ S300$& ..  & ..  & 4.3 & 0.01 /1E-04 /4.18 & 0.96 /0.15 /1.56 & 0.02 /0.85 /0.52 & 1.6E+04 & 0.4 & 0.21  &0.44 & Y\\\hline
$\bar{n}0.035\_ S2$& 585.6 &  4.58 & 164.3 & 0.56 /0.007 /2.82 & 0.44 /0.49 /0.93 & 0.004 /0.50 /0.55 & 3.5E+02 & 0.2 & 0.23 &1.51  & m\\
$\bar{n}0.035\_ S10$& ..  & ..  & 32.9 & 0.18 /0.002 /2.84 & 0.81 /0.27 /1.03 & 0.01 /0.73 /0.47 & 1.4E+03 & 0.3 & 0.18 &0.89 & Y \\
$\bar{n}0.035\_ S100$& ..  & ..  & 3.0 & -- /-- /-- & 0.96 /0.20 /1.95 & 0.04 /0.80 /0.69 & 6.1E+03 & 0.3 & 0.31 &0.60 & Y\\ \hline

  \end{tabular}
\footnotetext{$^1$TR $=$ Thermal Runaway, Y $=$ Yes, N $=$ No, m $=$ metastable state}
  \label{t:simIII}
\end{center}
\end{sidewaystable*} 

\subsection{Results}

\subsubsection{Slices of some output ISM}

\begin{sidewaysfigure*}
\begin{center} 
\includegraphics[width=1.1\textwidth]{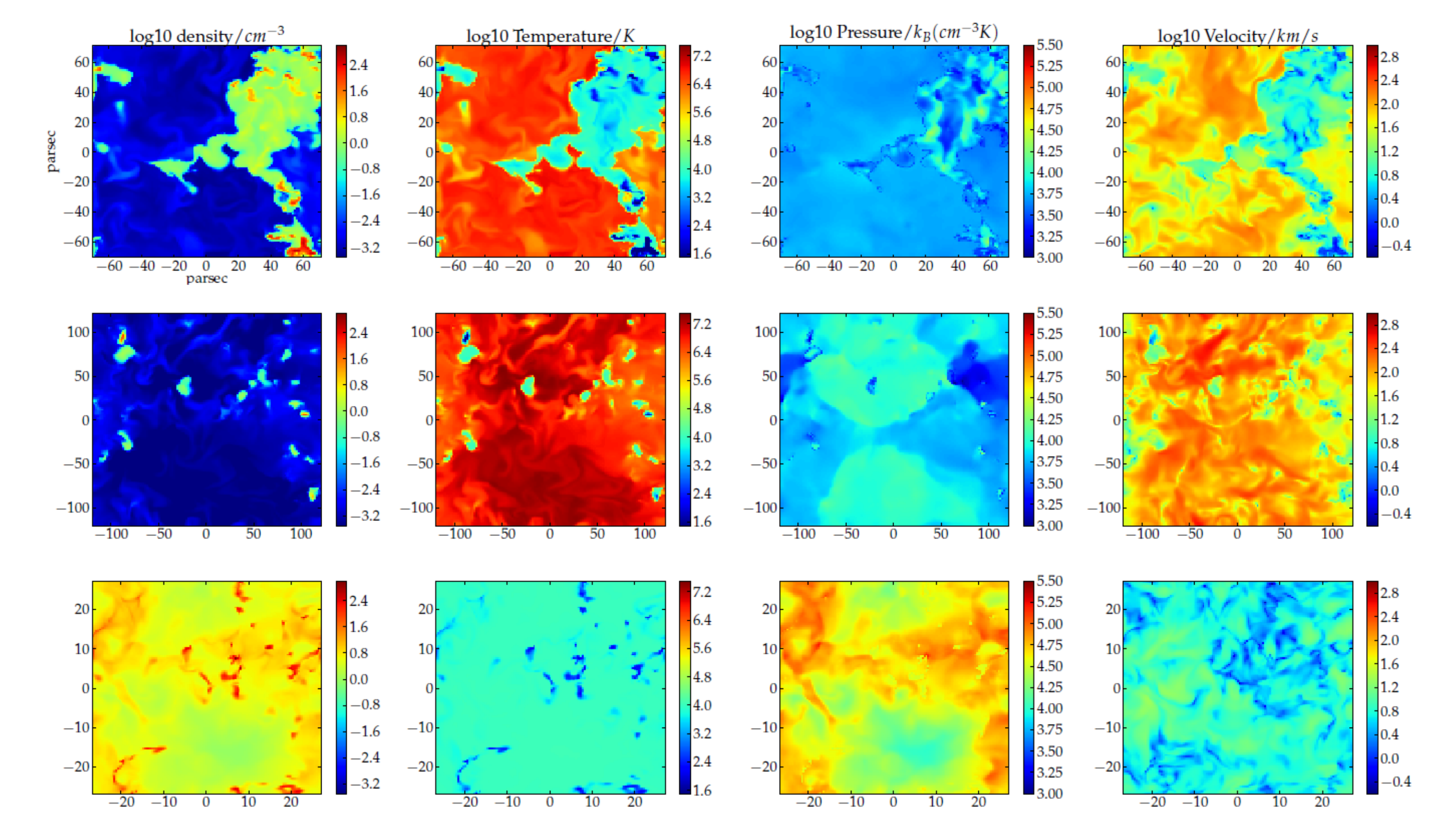}
\caption{Slices of density, temperature, pressure and velocity of the ISM from simulations with varying mean density $\bar{n}$ and SN rate $S$. Upper row: $\bar{n}1\_ S200$ (solar neighbourhood) at $t=104$ Myr; middle: $\bar{n}0.3\_ S100$ at $t = 46$ Myr; lower: $\bar{n}10\_ S3000$ at $t= 135$ Myr. Note that the box sizes vary. See Section \ref{IIImodel} for model details. Most volume is shared by the three typical phases with $T \sim 10^6, 10^4, 10^2$ K, respectively. Depending on $\bar{n}$ and $S$, a certain phase can be missing. The clumping of gas varies with $\bar{n}$ and $S$.
} 
\label{f:2d_all}
\end{center}
\end{sidewaysfigure*}

Fig. \ref{f:2d_all} shows slices of density, temperature, pressure and velocity of three simulation examples. The upper panels are for the run $\bar{n}1\_ S200$, corresponding approximately to the solar neighbourhood. Gas settles into three distinct phases with contrasting density and temperature: Hot $T\sim 10^{6.6}$ K, warm $T \sim 10^4$ K, and cold $T \sim$ a few hundred K, which have been observed by a variety of techniques and summarized in \cite{mckee77}. These phases are in rough pressure equilibrium, with variations of a factor of $\sim 30$, significantly smaller than the dynamical range of density and temperature. Velocity fields indicate that the ISM is turbulent, with the lighter component moving faster. Hot gas fills about 60\% of the volume after reaching steady state. These results agree with the previous analytical and and numerical simulations \citep[e.g.][]{mckee77, joung06,gent13,gatto15}.

Middle panels are for $\bar{n}0.3\_S100$, an example of a HDMM. SNRs overlap with each each other, rendering a connecting topology:  $\sim$ 90\% of the volume is filled with hot gas, and dense clouds are bathed in it.
The clouds mainly consist of the warm phase. The deficiency of the cold medium is attributed to the less efficient cooling of less dense ISM, and the PEH is able to keep the temperature of most clouds to $\sim 10^4$ K.

Lower panels show the results of $\bar{n}10\_S{3000}$, corresponding to a star forming region. The SN rate is insufficient to retain a hot phase. Hot bubbles close up before they overlap with each other. SN do not play a significant role in shaping the thermal state of the ISM. The two-phase medium is maintained by the balance between cooling and PEH. On the other hand, the momentum input from SN drives turbulent motions in the gas, and is therefore important for shaping the dynamical state of the ISM, which may have strong implications for star formation \citep{maclow04,mckee07}.

\subsubsection{Evolution of the different phases}

(1) $\bar{n}1\_ S200$ (MW disk-average): example of reaching steady state and pressure equilibrium 

\begin{figure}
\begin{center}
\includegraphics[width=1.1\textwidth]{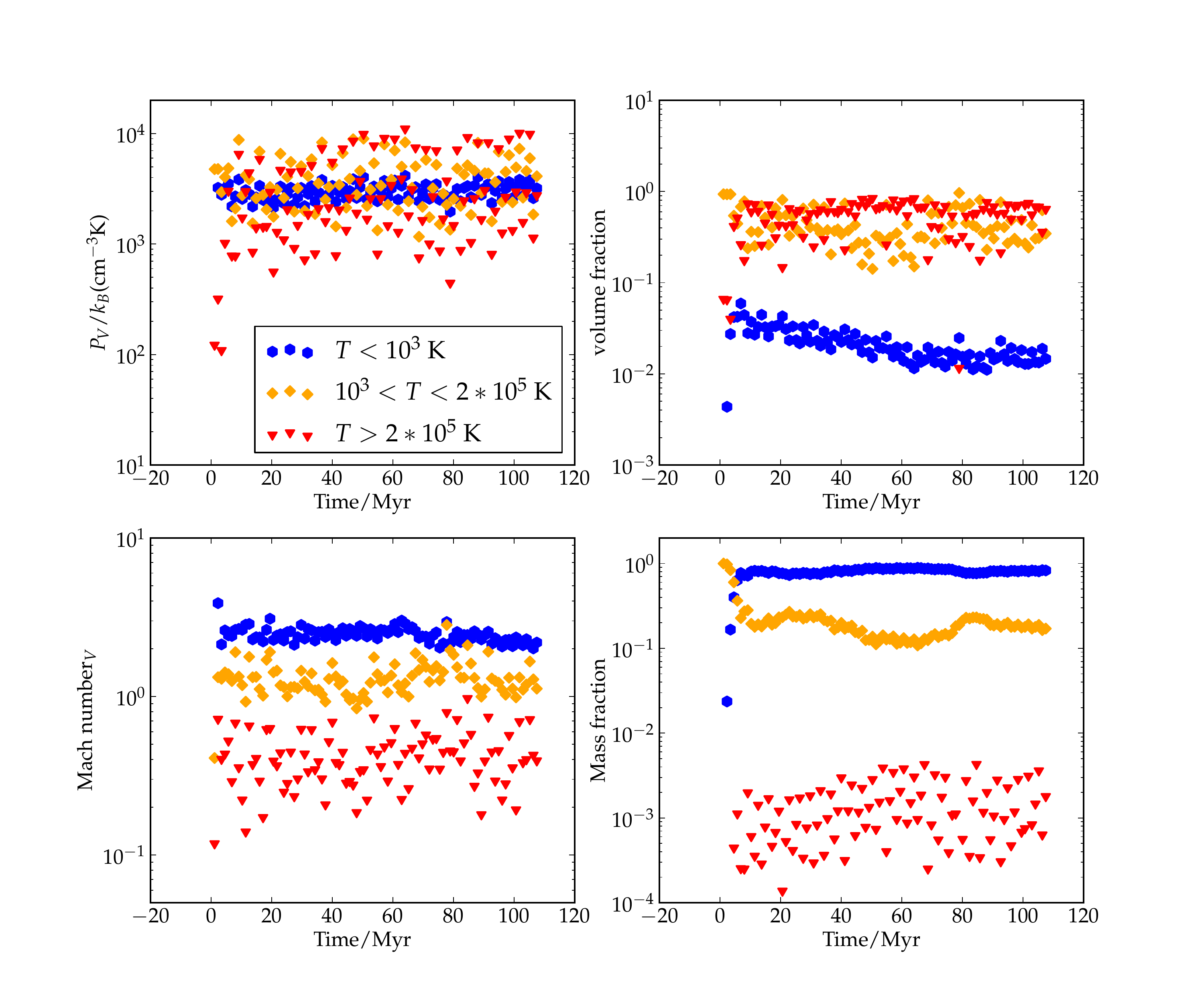}
\caption{Evolution of the three ISM phases for $\bar{n}1\_ S200$. 
Up to down, left to right: volume-weighted pressure $P_V/k_B$, Mach number, volume fraction, mass fraction. 
Red triangles, yellow diamonds and blue circles denote the cold ($T<10^3$ K), warm ($10^3$ K $< T < 2\times 10^5$ K ) and hot ($T>2\times 10^5$ K) phases, respectively. The medium reaches a steady state and the three phases are in rough pressure equilibrium. }
\label{f:3phase_n1_S200}
\end{center}
\end{figure}

Fig. \ref{f:3phase_n1_S200} presents the time evolution of the three phases for the run $\bar{n}1\_ S200$. We show in the four panels the volume-weighted pressure and Mach number, and the volume and mass fraction of each phase. Each data point represents one snapshot. The properties of the hot gas exhibit larger fluctuation, reflecting periodic explosions of SN. The evolution of the warm and cold phases, on the other hand, are smoother. Initially, the gas is uniformly 3000 K; within a few Myrs, cold clouds quickly build up and occupy $>80\%$ of the mass, and the cold and warm media both settle into their separate thermally-stable temperatures. Thereafter, the mass fraction and temperature for these two phases keep almost constant. The separation of the two phases happens as a result of the thermal instability, facilitated by perturbations from SN. The Mach number is higher for colder phases: the cold and warm media are mildly supersonic with Mach number $\sim 2.3$ and $ 1.3$, respectively, while the hot gas is subsonic (the slightly supersonic cold/warm phase and subsonic hot phase turn out to be the case for most of the runs, see Table \ref{t:simIII}).  All three phases are in approximate pressure equilibrium, with $nT \sim 3-4 \times 10^3 \cm-3$ K. These results agree very well with the observation \citep[e.g.][]{jenkins11}. Hot, warm and cold gas occupy 60\%, 40\%, 1.5\% of the volume, respectively, consistent with the observations in the solar neighbourhood \citep[e.g.][]{heiles03, murray15}. Note that observations of 21 cm absorption-line yield about 60\% of the mass of neutral hydrogen is warm \citep{heiles03}, which seems inconsistent with our result. However, we use $T<10^3$ K as the criterion for being ``cold'', different from the the observation, which is roughly $T\lesssim 200$ K. Also we do not distinguish molecular gas from the cold medium (nor do we track the ionization state) in our simulations. The mass ratio between cold HI and molecular is about 1.38 for the MW \citep{draine11}. Taking this into consideration, the revised mass ratio of cold to warm in our simulations becomes 3.3 (even assuming all warm gas is neutral), still larger than the observed 0.67. Some effects that we do not include may account for the over-condensation of the cold phase, for example, UV from massive stars can ionize and heat the cold medium; thermal conduction by hot gas may evaporate some of the cold gas and form a warm component. Another possibility is that if a few SN happen to have exploded in the dense clumps, they can effectively turn the cold medium into warm and hot \citep[e.g.][]{gatto15}. The stellar winds from their progenitors can have a similar effect. We will discuss more on the position of SN in Section 5.1.

(2) $\bar{n}0.1\_ S50$ (MW disk-halo interface): example of a HDMM and  ``thermal runaway"

The $(\bar{n},S)$ of this run corresponds to the disk-halo interface of the MW (see Fig. \ref{f:fvh} below). In the left panel of Fig 9, we show the volume-weighted pressure evolution for the three phases. Same as in Fig. 8,  each data point is calculated for one snapshot. Hot gas is over-pressured compared to the other two phases, which reflects what we have found in the previous experiment: the blast wave travels preferentially in the hot medium. It would take roughly the sound-crossing time for the clouds to re-establish the pressure equilibrium. The overlapping SN determine the overall properties of the ISM. The medium does not reach a steady state at the end of the simulation, and it is not obvious that it will ever reach one. This is because, despite the rising density of the clouds which enhances cooling, the volume is increasingly occupied by low-density hot gas where cooling is negligible; since blast waves lose the majority of their energy in the clouds, cooling is inevitably delayed until the shock sweeps enough volume. Similar phenomena have also been observed in some of the simulations in \cite{gatto15}. \cite{scannapieco12} discovered a transitional 1D velocity dispersion $\sim$ 35 km/s for the turbulent ISM, above which the rarefied gas cannot cool efficiently and the ISM undergoes a thermal runaway. We can understand the thermal runaway state in this way: if the SN rate is high enough for multiple SN to overlap, the ISM will be multiphse where the hot bubbles connect to each other. As we have found in the experiment in which a single SN propagates in a HDMM, the volume encompassed by the blast wave is much larger, and the cooling is suppressed, thus the spatial-temporal domain of a single SN ($V_{\rm{4D}}$) is greatly increased. This makes the overlapping of the SN reinforces itself, which leads the ISM to a ``runaway'' state.
In real galaxies, the persistent increase in pressure implies that the gas will expand, and hot gas can blow out and form winds (which are not captured in the periodic box simulations).

(3) $\bar{n}10\_ S3000$ (SF region): example of a subordinate hot phase

\begin{figure}[htb]
\centering
  \begin{tabular}{@{}cc@{}}
    \includegraphics[scale=0.55]{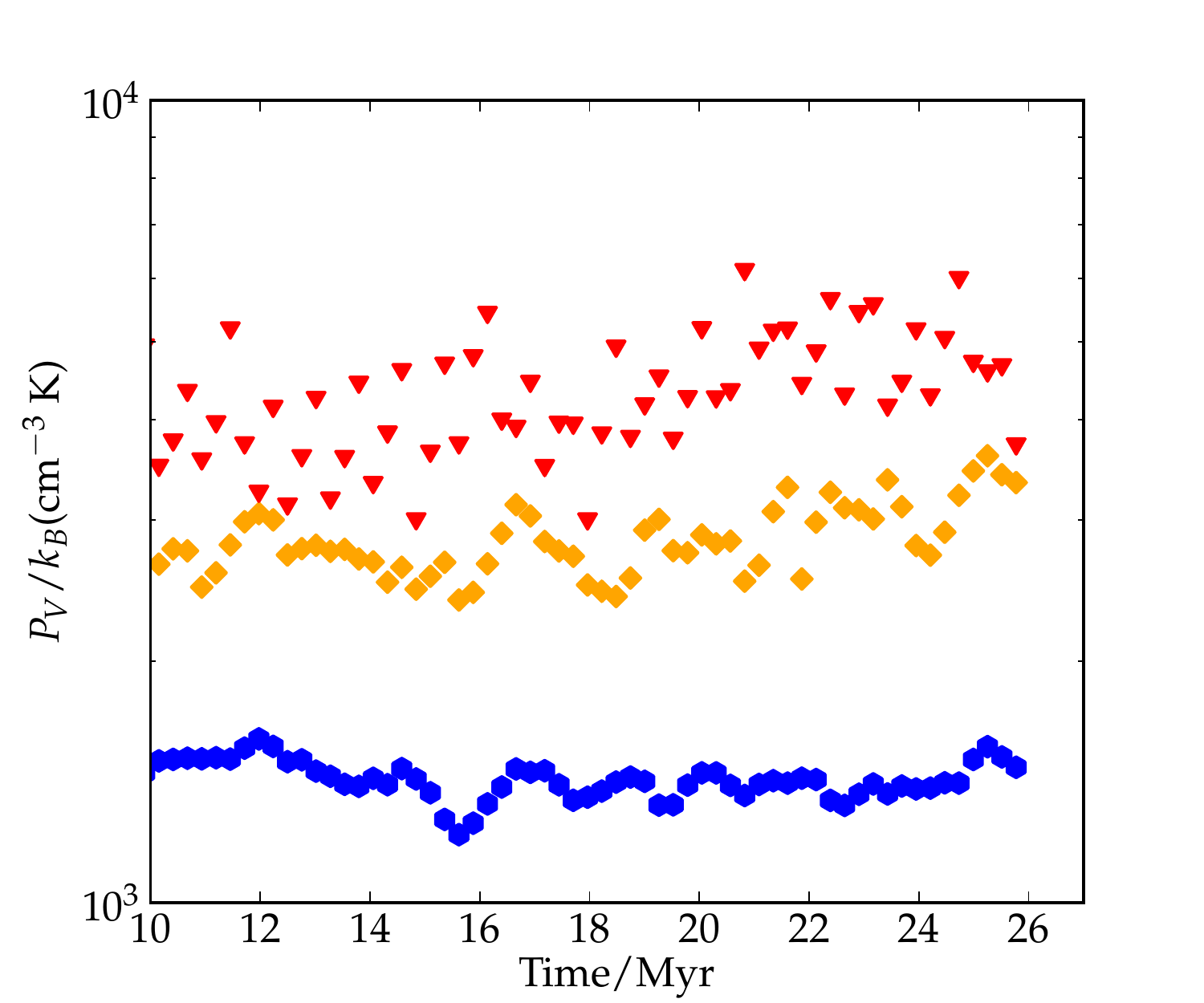} &
    \includegraphics[scale=0.55]{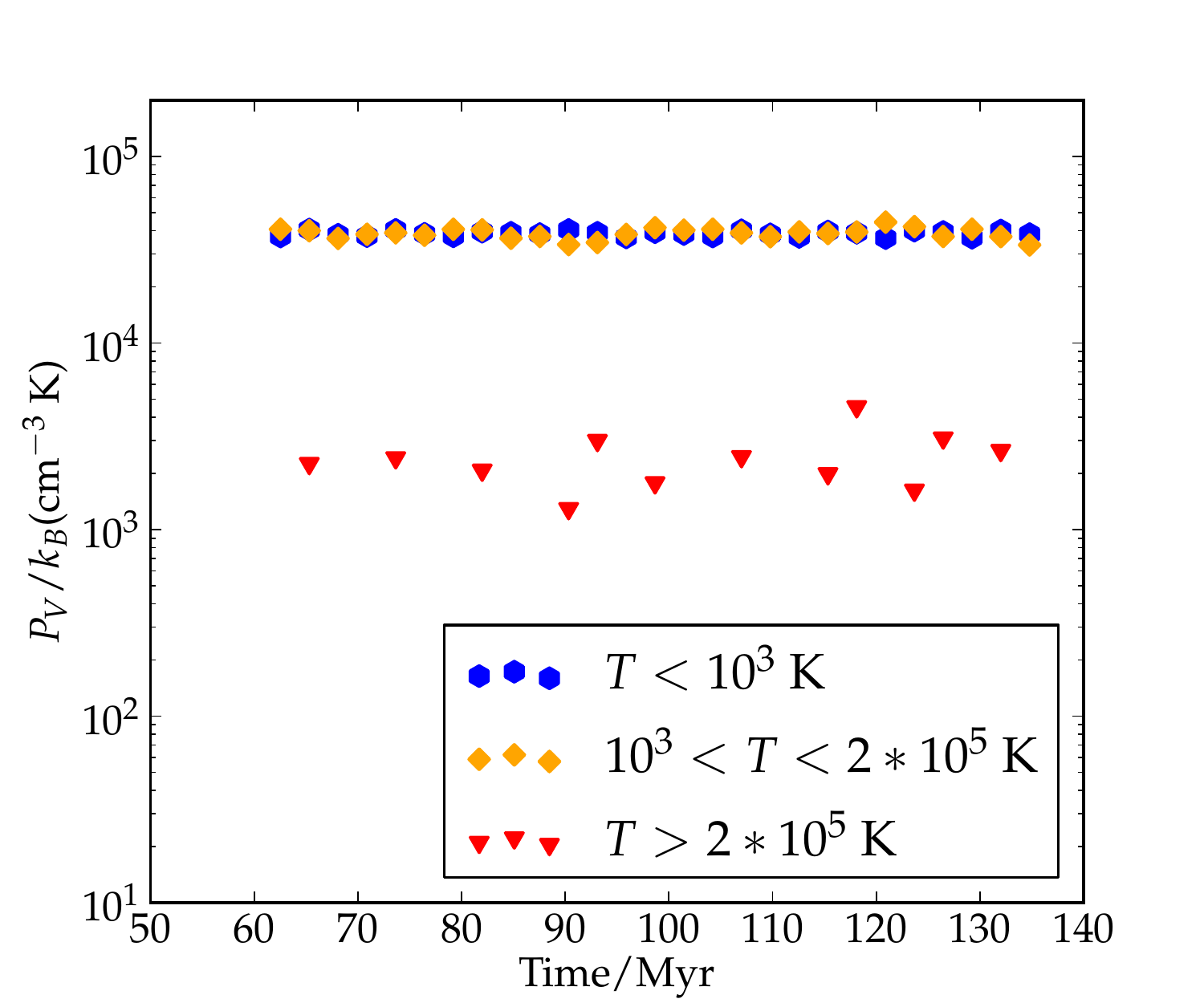}
  \end{tabular}
  \caption{Volume-weighted pressure $P_V$ of the three phases versus time for runs $\bar{n}0.1\_ S50$ (left) and $\bar{n}10\_ S3000$ (right). For $\bar{n}0.1\_ S50$,  the hot gas is over-pressured and the ISM is in a ``thermal runaway" state. For $\bar{n}10\_ S3000$, the hot phase is only present briefly after SN explosions and is significantly under-pressured. }
\label{f:P_t}  
\end{figure}

As shown in the right panel of Fig. \ref{f:P_t}, the ISM in this case has only warm and cold phases most of the time. Hot gas is present only briefly after SN explosions. SN are not determining the general thermal state of the ISM, since the pressure of the hot gas is significantly smaller than the other phases, and the time-averaged volume fraction is no more than a few percent (see Fig. \ref{f:fvh} below). The reason for the under-pressure of the hot phase is this: once the SNR enters the ``pressure-driven snowplow'' phase, hot gas moves much faster than the shell, and quickly runs into it and loses energy, which leaves the hot bubbles at an extremely low density, and therefore low pressure (Fig. \ref{f:single}). Note that right after SN explosions, the hot gas is over-pressured, but it only lasts till the shock becomes radiative, and the time interval is very short compared to our sampling. In fact, according to Eq. \ref{eq:tcool},  $t_{\rm{cool}} =$ 0.014 Myr for $n = 10\cm-3$, and yet the time interval between two consecutive SN is 2.08 Myr, and our sampling will not catch the over-pressured stage. In other words, if we could sample the data at a frequency larger than every $10^4$ year, we would see the very spiky evolution of the hot gas pressure; if we could enlarge our box size (but keep $S$ unchanged) so that at any given time, there are many SN exemplifying different stages of their evolution, we would see the mean pressure of hot gas go beyond that of warm/cold, because the spatially-averaging is biased towards regions with pressure that is orders-of-magnitude higher than the median. In other words, when we say hot gas pressure is lower than that of the other phases, we mean that at any given time, if we randomly choose a location that is above $2\times 10^5$ K, it is most likely to be under-pressured.

(4) \fvh$\sim 0.6\pm 0.1$: metastable state of the ISM and transition to ``thermal runaway"

\begin{figure}
\begin{center}
\includegraphics[width=1.1\textwidth]{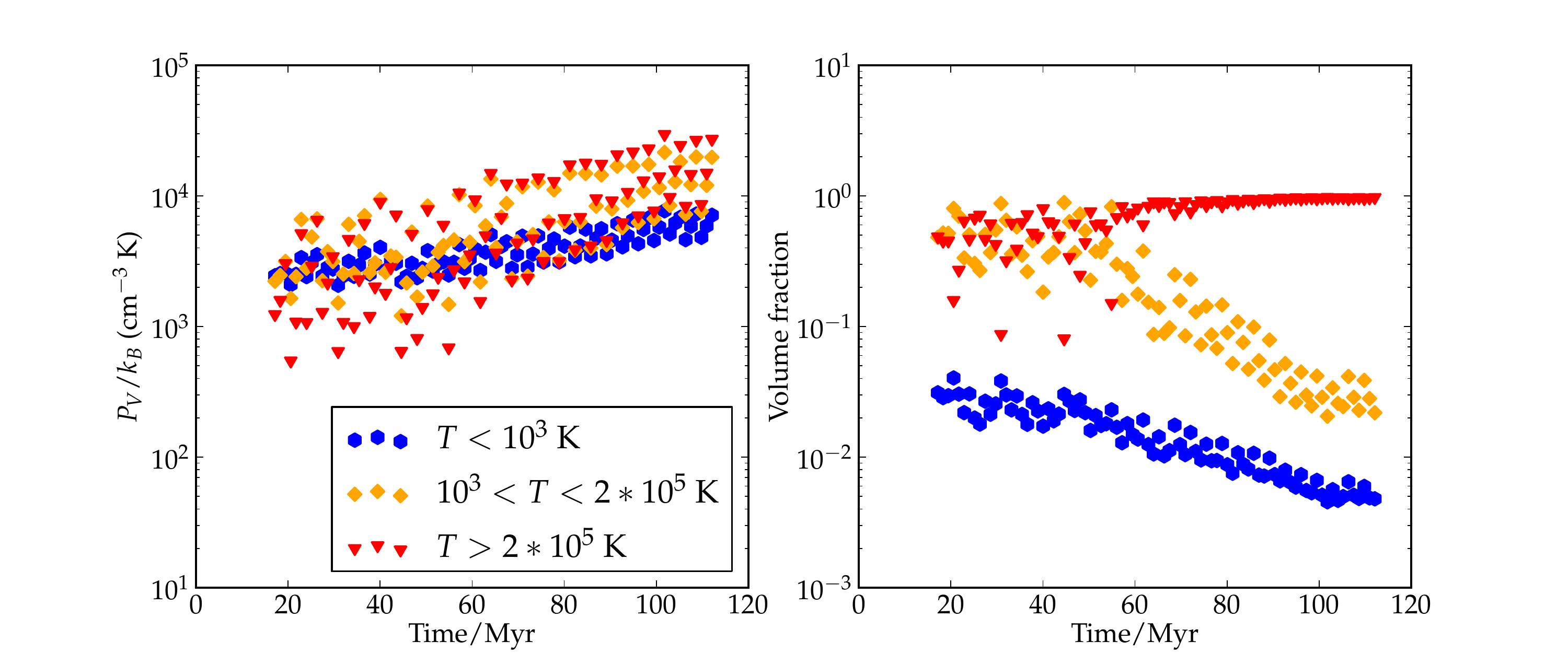}
\caption{Transition from a metastable state to ``thermal runaway" for $\bar{n}1\_ S200$. Initially, the ISM reaches the apparent steady state, when roughly half the volume is hot.  The transition happens at $t\sim 55$ Myr, when the hot gas quickly dominates pressure and $> 90\% $ of the volume, and pressure of all phases rises progressively.}
\label{f:3phase_n1_S200_transition}
\end{center}
\end{figure}

For the simulations which reach a steady state and have $f_{\rm{V,hot}} \sim 0.5-0.7$, we sometimes observe an interesting transition from ``steady state" to ``thermal runaway". Fig. \ref{f:3phase_n1_S200_transition} exemplifies such a transition. This is for $(\bar{n},S)=(1,200)$, with the same set-ups as the fiducial run. The locations of the SN are random, but different from the fiducial one. Initially the ISM reaches a steady state (same as Fig. \ref{f:3phase_n1_S200}): Each of the three phases occupies a stable portion of the volume, and they are roughly in pressure equilibrium. Starting from $t\sim 55$ Myr, however, hot gas quickly comes to dominate the space, while the volume fraction of the warm phase drops from $\sim 50\% $ to a few percent. Meanwhile, the gas pressure rises progressively, and the equilibrium is disrupted -- the hotter phase has a higher pressure. These phenomena are the same as those in a ``thermal runaway" state. The transitions are somewhat stochastic, and can happen at any time after the steady state is established (the example shown in Fig. \ref{f:3phase_n1_S200} is stable to the end of the run $t=110$ Myr). The stochastic transition is not so surprising: the configuration of the ISM with half the volume hot is on the verge of changing from ``isolated bubbles'' to a ``connecting'' topology; any new SN may break the balance and lead to the topology transition. Once the hot bubbles connect, any new SN will have a much larger $V_{\rm{4D}}$, and the overlapping will inevitably drive the ISM toward ``thermal runaway".  Therefore, a steady state with $f_{\rm{V,hot}} \sim 0.6\pm 0.1$ is better termed as ``metastable" due to its fragility.

\subsubsection{$f_{\rm{V,hot}}$ as a function of $(\bar{n},S)$}

\begin{figure*}
\begin{center} 
\includegraphics[width=1.3\textwidth]{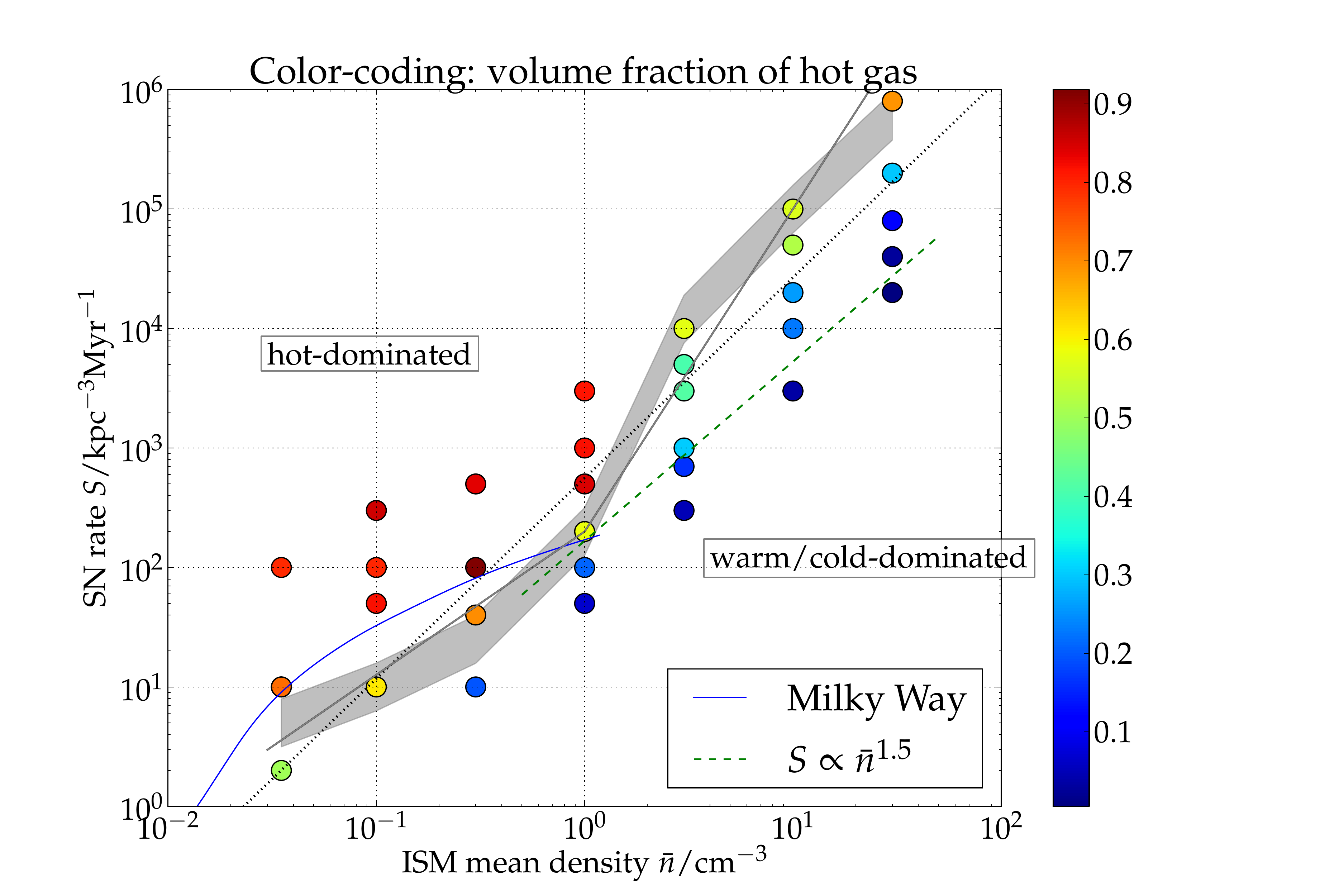}
\caption{Volume fraction of the hot ($T>2\times 10^{5}$ K) gas $f_{\rm{V,hot}}$ as a function of the mean gas density $\bar{n}$ and SN rate $S$. Each circle represents one simulation. Colors show the \fvh. The blue solid line indicates $S$ as a function of $\bar{n}$ from MW mid-plane to halo (see Section \ref{sec:para} for a description). The green dashed line indicates $S\propto \bar{n}^{1.5}$. $(\bar{n},S)=(1,200)$ is roughly for the MW-average. The shaded area shows the transitional region, i.e. $f_{\rm{V,hot}} \approx 0.6 \pm 0.1$ -- above it the hot gas dominates the volume and the ISM undergoes thermal runaway, and below it the SN play a subordinate role and most volume is warm/cold; within the transitional region the ISM is metastable. The gray solid line indicates the nominal line to roughly represent $S_{\rm{crit}}$ above which thermal runaway happens (Eq. \ref{eq:fvhot}). The black dotted line shows \fvh$=0.6$ from the simple analytical expectation Eq. \ref{eq:fvhot1}.  }
\label{f:fvh}
\end{center}
\end{figure*}

Fig. \ref{f:fvh} shows how \fvh of the ISM changes when we systematically vary $\bar{n}$ and $S$. Each simulation is represented by one circle on the $\bar{n}-S$ diagram, and the colors indicate \fvh. Each data point is obtained by averaging over 20 snapshots evenly sampled during the last $20\%$ of the simulation time. From the figure, a lower $\bar{n}$ and higher $S$ yield a higher $f_{\rm{V,hot}}$. As mentioned above, those with $f_{\rm{V,hot}} \approx 0.6 \pm 0.1$ lie in the ``transitional region" (grey-shaded), where the ISM is metastable and can change into ``thermal runaway" with any new SN. Below the ``transitional stripe", SN do not play a major role in shaping the ISM, and can be significantly under-pressured; above that, the hot gas has a ``connecting'' topology and is in ``thermal runaway" state, with hot gas dominating the volume and an increasing pressure.

A simplest estimate of the \fvh is 
\begin{equation}
f_{\rm{V,hot}} \sim  \frac{4\pi}{3} (\phi R_{\rm{cool}})^3 t_{\rm{close}}S. 
\label{eq:fvhot}
\end{equation}
$\phi R_{\rm{cool}}$ is the size of the hot bubble, where $\phi$ is about 1.6-1.8 (Section \ref{sec:I}). $t_{\rm{close}}$ is the time for a SNR to close up: 
\begin{equation}
 t_{\rm{close}} \sim \phi R_{\rm{cool}}/c_s.
\label{eq:t_close}
\end{equation}
Therefore,
\begin{equation}
f_{\rm{V,hot}}
 \sim 0.11(\frac{\phi}{1.7})^4 (\frac{10 \rm{km/s}}{c_s}) (\frac{S}{100 \kpc^{-3} \myr-1}) (\frac{\bar{n}}{1 \cm-3})^{-1.68} (\frac{E_{\rm{SN}}}{10^{51}\erg})^{1.16}.
\label{eq:fvhot1}
\end{equation}

Eq. \ref{eq:fvhot}-\ref{eq:fvhot1} only hold true when $f_{\rm{V,hot}} \lesssim 0.6$, i.e. SNRs do not overlap and thus thermal runaway does not come into operation. According to Eq. \ref{eq:fvhot1}, given an $\bar{n}$, $f_{\rm{V,hot}}$ should be proportional to $S$. This seems to be (very roughly) the case for $\bar{n}1\_S50$, $\bar{n}1\_S100$, $\bar{n}1\_S200$. However, the simulations show that for a higher density medium, the dependence of $f_{\rm{V,hot}}$ on $S$ is sublinear. For example, when $\bar{n} = 3\cm-3$, a factor of 10 increase of $S$ -- from $10^3$ to $10^4\kpc^{-3} \myr-1 $-- only leads to a factor of $\sim 2$ increase in $f_{\rm{V,hot}}$.  Another way to look at it is to plot the corresponding curve of $f_{\rm{V,hot}} = 0.6$ using Eq. \ref{eq:fvhot1} (black dotted line on Fig. \ref{f:fvh}). For $\bar{n}>1\cm-3$, the simulations with$f_{\rm{V,hot}} \approx 0.6\pm 0.1$ (grey-shaded) have an $S$ above the simple theoretical expectation at a given $\bar{n}$; whereas for $\bar{n}\lesssim 1\cm-3$, the critical $S$ to achieve a HDMM has a weaker dependence on $n$ than expected. Overall, thermal runaway appears harder to achieve in dense gas.

We argue that one key factor to explain the discrepancy is the photoelectric heating. Recall that we have assumed $\Gamma_{\rm{pe}} \propto S$. If we consider a medium without SN (but with normal cooling and PEH), then the warm and cold phases would co-exist with the same pressure, each in its thermally stable state. If the mean density of the ISM is fixed, a higher $\Gamma_{\rm{pe}}$ means more mass in the warm phase. Note that the cold medium usually occupies a small volume whereas the warm phase permeate the space, so more mass in the warm phase suggests that a SNR will travel effectively in a ``denser" medium, in which it would have a smaller $R_{\rm{cool}}$. This would have a substantial negative effect on $f_{\rm{V,hot}}$ when we consider multiple SNRs, since $f_{\rm{V,hot}}$ depends on $R_{\rm{cool}}$ sensitively as $ f_{\rm{V,hot}} \propto R_{\rm{cool}}^4\propto n^{-1.68}$ (Eq. \ref{eq:fvhot} and \ref{eq:t_close}). 

\begin{sidewaysfigure}[ht]
\begin{center} 
\includegraphics[width=1.2\textwidth]{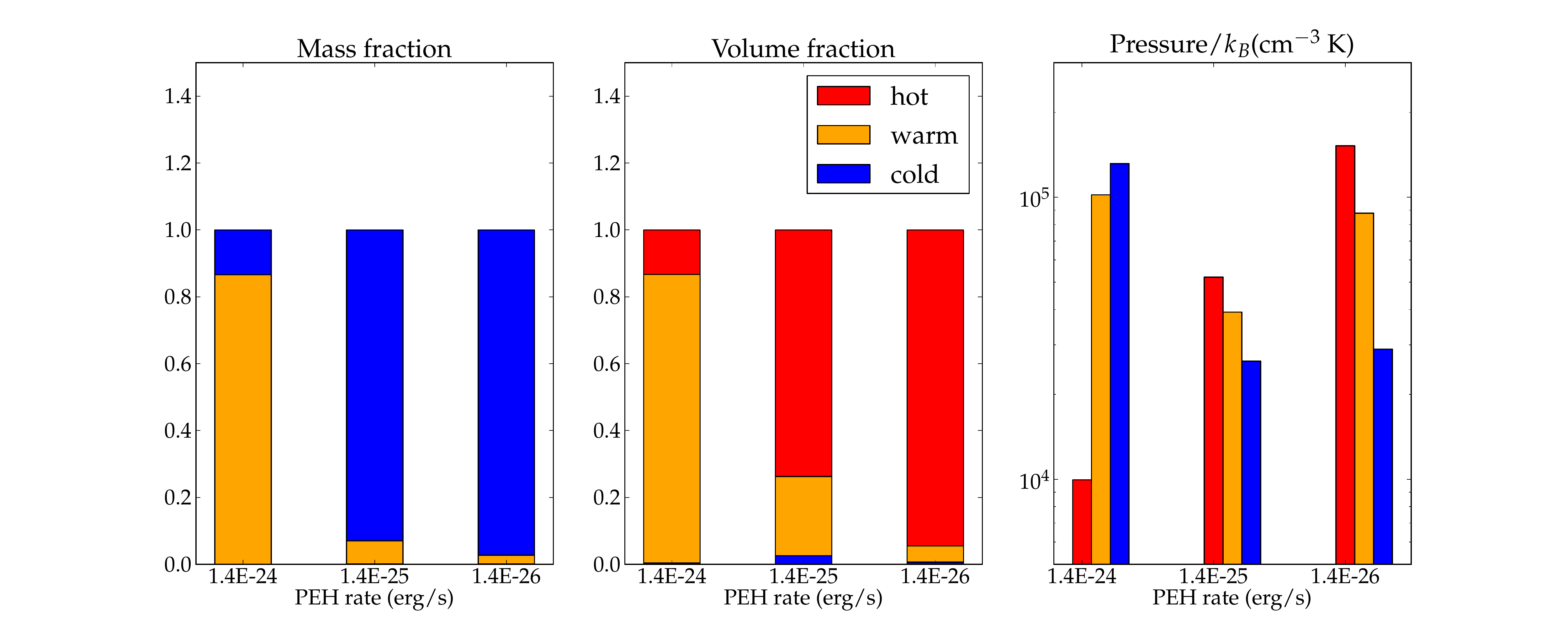}
\caption{Comparison of the ISM from the run $\bar{n}10\_ S1000$ with different photoelectric heating rates. Colors represent the three ISM phases. }
\label{f:pe_rate}
\end{center}
\end{sidewaysfigure}

To test this idea, we take the run $\bar{n}10\_S10000$, and reduce its original PEH rate $\Gamma_{\rm{pe}}$  by a factor of 10 and 100, respectively. The latter case corresponds to a PEH rate similar to the solar neighbourhood. The comparison is shown in Fig. \ref{f:pe_rate}. 

The effect of $\Gamma_{\rm{pe}}$ is obvious. Hot gas occupies more volume and has higher pressure for lower PEH rates. For the lowest PEH run, the ISM is clearly in a thermal runaway state, since the hot gas occupies most volume and determines the pressure. By "determining the pressure" we mean that, if no SN explodes, the pressure of a thermally stable two-phase medium would be about $3\times 10^3\cm-3$K for a PEH rate of $1.4\times 10^{-26}$ erg/s, but now the ISM pressure is 1-2 orders of magnitude higher than the equilibrium pressure. In this case the overlapping SNRs dominates the thermal state of the ISM. The mass fraction of the cold gas is also much higher in the low PEH run, which is likely due to two effects: a smaller background heating rate that allows more gas in the cold phase, and compression from the overlapping SNRs. Interestingly, the volume-weighted pressure does not have a monotonic dependence on $\Gamma_{\rm{pe}}$: the intermediate $\Gamma_{\rm{pe}}$ case has the lowest pressure among the three. This exhibits a subtle balance between the hot and warm phases:  SN marginally overlap but are able to get rid of most of the energy; the pressure is slightly enhanced from that of a two-phase medium. PEH plays the leading role in determining the ISM properties in the highest $\Gamma_{\rm{pe}}$ case, whereas SN wins in the lowest one. Note that in Section \ref{sec:I}, we have shown that PEH rate does not impact the SN evolution in a uniform medium. It affects the ISM through regulating the warm/cold ratio in the gas reservoir. We emphasize that the PEH is a critical process in the ISM -- it plays an important role in determining the portion of various phases and the overall pressure \citep{wolfire95, wolfire03}. This may also have strong implications for the global structure of the disk and star formation \citep{tasker11}.

We give a simple fitting formula for the critical SN rate $S_{\rm{crit}}$ as a function of n, above which thermal runaway occurs, i.e. \fvh$(\bar{n}, S_{\rm{crit}})\approx 0.6$:
\begin{equation}
S_{\rm{crit}} =  200 (\frac{\bar{n}}{1\cm-3})^k (\frac{E_{\rm{SN}}}{10^{51}\erg})^{-1} \kpc^{-3} \myr-1, 
\label{eq:S_crit}
\end{equation}
where $k = (1.2,2.7)$ for $\bar{n} < 1,\ >1 \cm-3$, respectively. We have shown this with a grey solid line in Fig. \ref{f:fvh}.

\subsubsection{Pressure as a function of $(\bar{n},S)$}

\begin{figure*}
\begin{center} 
\includegraphics[width=1.3\textwidth]{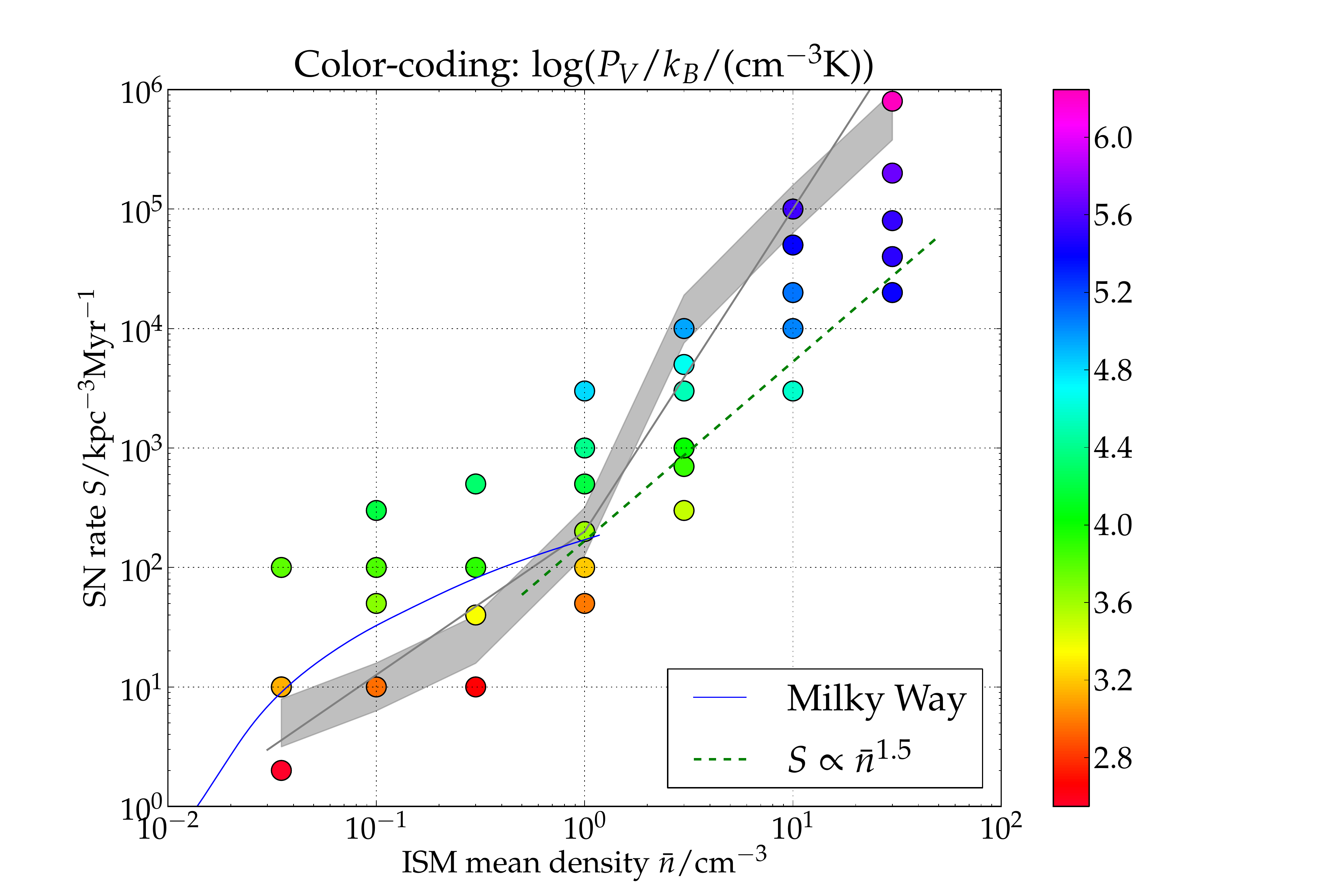}
\caption{ Same as Fig. \ref{f:fvh}, but colors show the volume-weighted pressure $P_{\rm{V}}/k_{\rm{B}}$.  }
\label{f:Pv}
\end{center}
\end{figure*}

Fig. \ref{f:Pv} is similar to Fig. \ref{f:fvh}, but the color-coding indicates the volume-weighted pressure $P_{V}/k_B$ for each simulation. The pattern of $P(\bar{n},S)$ is clear. For a fixed $\bar{n}$, pressure increases with $S$, which is expected since higher S means higher heating rate. The scaling is nearly linear. The dependence of $P_V$ on $\bar{n}$ is less sensitive, and the correlation is mostly negative, at least for $S\lesssim 10^4\kpc^{-3}\myr-1$.

\begin{figure*}
\begin{center} 
\includegraphics[width=1.1\textwidth]{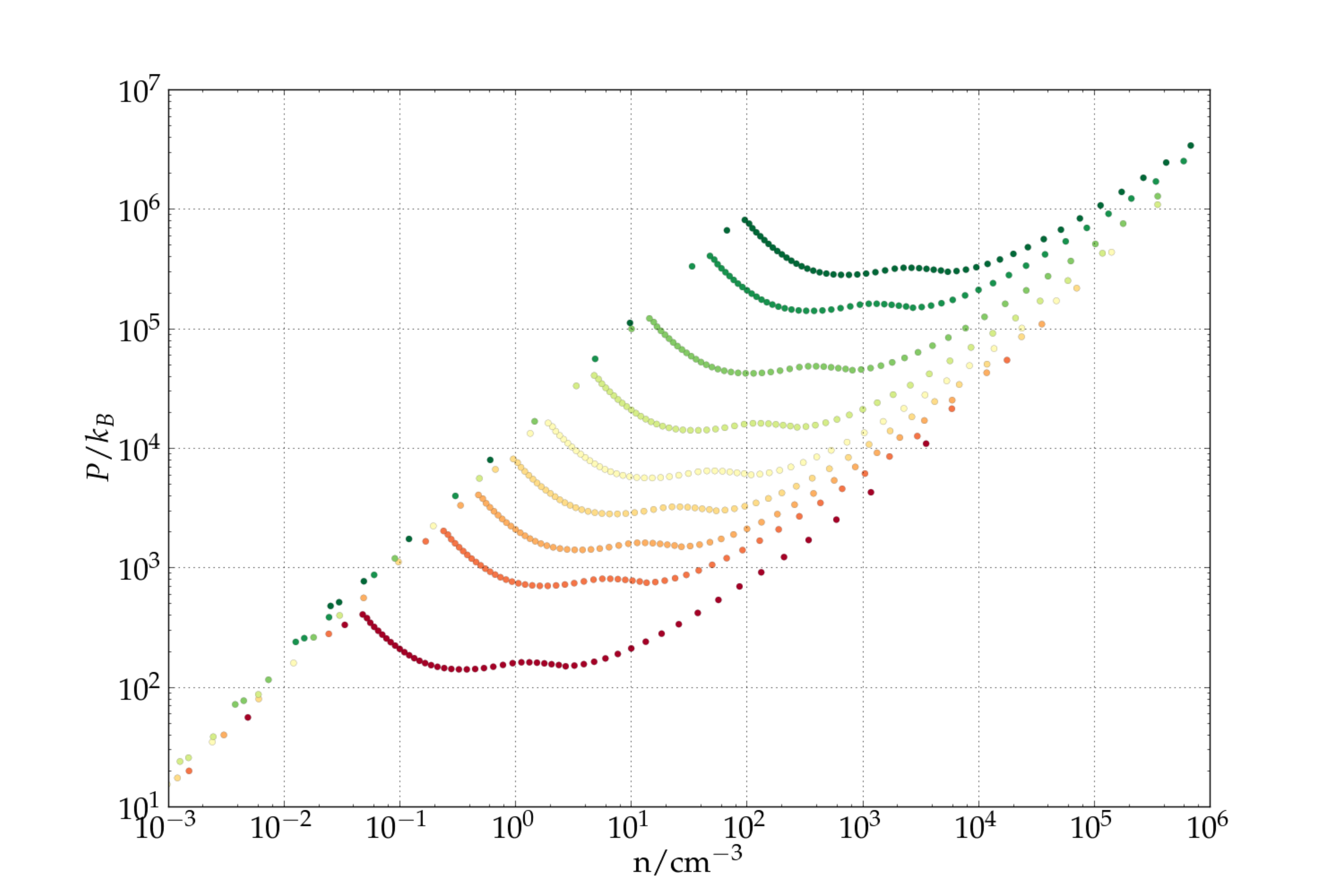}
\caption{Equilibrium pressure as a function of density $n$ for $T\lesssim 10^4$ K gas for a variety of photoelectric heating rates. Red to green: 0.1, 0.5, 1, 2, 4, 10, 30, 100, 200 $\times$ 1.4E-26 erg/s per H atom.  The calculation is based on the cooling curve adopted in our simulation.}
\label{f:P_n}
\end{center}
\end{figure*}

Can we understand $P_V$ as a function of $(\bar{n},S)$ more quantitatively? We have seen that it is not easy to obtain a very accurate \fvh from a simple analytic argument, especially for the thermal runaway case when \fvh keeps rising. On the other hand, if \fvh is known, can we predict pressure? In fact, if we neglect SN and only consider a two-phase medium, one can predict quantities such as temperature, pressure, density of each phase for a thermal equilibrium condition, given a cooling curve and a heating term \citep[e.g.][]{draine78} (note that under some circumstances, there may only be one phase). If the two phases coexist, they will reach pressure equilibrium. The equilibrium solution can be shown  by a phase diagram, which is the correlation between any two of the three thermodynamic variables, density, temperature and pressure. We show in Fig. \ref{f:P_n} examples of the equilibrium $P_{\rm{eq}}-n$ relation, given our cooling curve and a variety of PEH rates. The equilibrium solution is thermally stable for the part of the curve with positive slope. The impact of different PEH rates on the equilibrium pressure of the neutral phase has been explored by \cite{wolfire95, wolfire03}. Our simple prescription predicts a very similar $P_{\rm{eq}}-n$ relation to their more detailed model. Take the curve with the MW PEH rate for an instance (the third curve from the bottom): the unstable solution corresponds to roughly in between 0.4-15 $\cm-3$, and most part outside is stable (except a second dip between $15-30\cm-3$). When the two phases coexist, the warm phase has $n_{\rm{w,eq}} \approx 0.4\cm-3$ and $T\approx 10^4$ K, and the cold has $n_{\rm{c,eq}} \approx 15\cm-3$ and $T\approx 250$ K. Another way to look at this relation is that, if a mean gas density $\bar{n}$ is in the range [$n_{\rm{w,eq}},n_{\rm{c,eq}}$], the medium will settle into two phases with pressure $P_{\rm{eq}}/k_B \approx 3\times 10^3 \cm-3$ K; if $\bar{n}< n_{\rm{w,eq}}$, the medium will be all warm and $P \propto \bar{n}$; if $\bar{n}> n_{\rm{c,eq}}$, the gas is all cold with roughly $P\propto \bar{n}^{0.5}$. 
 
When we consider SN and the hot phase, the hot component can occupy a significant volume but usually contains negligible mass, so the mean density for the warm/cold phase is simply $\bar{n}/(1-f_{\rm{V,hot}})$. Therefore, given an $n$, PEH rate and $f_{\rm{V,hot}}$, the pressure can be predicted in the following way. The equilibrium densities of warm and cold phase scale linearly with PEH:
\begin{equation}
n_{\rm{w,eq}} \approx  0.4\cm-3 (\frac{\Gamma_{\rm{pe}}}{1.4\times 10^{-26} \rm{erg/s}});\ \ \ \ \\
n_{\rm{c,eq}} \approx 15\cm-3 (\frac{\Gamma_{\rm{pe}}}{1.4\times 10^{-26} \rm{erg/s}}).
\label{eq:neq}
\end{equation}
Then the pressure is (assuming the warm phase has a constant temperature $10^4$ K):
\begin{equation}
P_{\rm{V,pred}}/k_B= \begin{cases}
\frac{\bar{n}}{1-f_{\rm{V,hot}}}\times 10^4 \rm{K}  & \text{if $\frac{\bar{n}}{1-f_{\rm{V,hot}}}\le n_{\rm{w,eq}}$ },  \\
n_{\rm{w,eq}}\times 10^4\rm{K} & \text{if $n_{\rm{w,eq}} <\frac{\bar{n}}{1-f_{\rm{V,hot}}}< n_{\rm{c,eq}}$}, \\
n_{\rm{w,eq}}\times 10^4\rm{K} (\frac{\bar{n}}{1-f_{\rm{V,hot}}}/n_{\rm{c,eq}})^{0.5} & \text{if $\frac{\bar{n}}{1-f_{\rm{V,hot}}}\ge n_{\rm{c,eq}}$}.
\end{cases}
\label{eq:P_pred}
\end{equation}
In Table \ref{t:simIII} we show the ratio between our analytic calculation and the simulation results  $P_{\rm{V,pred}}/P_V$. They agree within a factor of 3 generally. Note that in the thermal runaway case, the hot gas pressure is higher than that in the warm and cold, but the ratio is on the order of unity.

In cosmological simulations where the resolution in space and time is low, the multi-phase structure of the ISM and individual SN are often not resolved, and the collective SN feedback may be modelled as a sub-grid effective equation of state, if the the cell size is much larger than the spatial domain of a single SN, and the timestep is longer than $t_b$ (Eq. \ref{eq:tb}). Although Eq. \ref{eq:P_pred} can predict the pressure, it has an extra parameter $f_{\rm{V,hot}}$ which cannot be easily obtained. We are then tempted to fit $P_V$ directly as a function of $\bar{n}$ and $S$. An advantage is that from Fig. \ref{f:Pv}, the pattern of $P_{V}(\bar{n},S)$ seems clear and simple and power-law-like. We thus fit the output data using the least squares method, which yields:
\begin{equation}
P_V/k_B = 10^{3.47 \pm 0.05}  \cm-3 \rm{K}  (\frac{S}{100 \kpc^{-3} \myr-1 })^{0.87 \pm 0.05} (\frac{\bar{n}}{1\cm-3})^{-0.33 \pm 0.07}.
\label{eq:P}
\end{equation} 
The indices show that $P_V$ scales almost linearly with $S$, while it is somewhat less dependant on $\bar{n}$. The fitting formula gives a reasonable prediction of pressure for $\bar{n}< 30\cm-3$. We caution that the above fitting formula is time-independent. It is fine for an ISM that reaches equilibrium, but for those thermal runaway ones, the pressure is not constant but is still rising at the end of the simulation. To generalize the result to the thermal runaway cases, we extend Eq. \ref{eq:P} by adding a factor of $t/t_{\rm{sim}}$ to allow for time evolution. Just to remind readers, $t_{\rm{sim}}=2.5t_b$, and $t_b$ is given in Eq. \ref{eq:tb} as a function of $(\bar{n},S)$. Thus, 
\begin{equation}
P_V/k_B= \begin{cases}
2.9\times 10^3 \cm-3\rm{K} (\frac{S}{100 \kpc^{-3} \myr-1 })^{0.87} (\frac{\bar{n}}{1\cm-3})^{-0.33}\  &\text{if $S<S_{\rm{crit}}$, } \\
9.6\times 10^3 \cm-3\rm{K} (\frac{S}{1000 \kpc^{-3} \myr-1 })^{1.87} (\frac{\bar{n}}{1\cm-3})^{-1.59} (\frac{t}{10\rm{Myr}})\  &\text{if $S\geqslant S_{\rm{crit}}$. } \\
\end{cases}
\label{eq:P_fit}
\end{equation}
$S_{\rm{crit}}$ is given in Eq. \ref{eq:S_crit}, which only depends on $\bar{n}$ and $E_{\rm{SN}}$. We emphasize that the above model holds only under the following conditions: (1) the spatial domain is at least a few times the size of a single SNR, and (2) the elapsed time $t$ is on the order of a few $t_b$. If $t\ll t_b$, the ISM is not fully impacted  by SN, and if $t \gg t_b$ and the ISM is in the thermal runaway state, the gas will probably expand due to the increasing pressure. In cosmological simulations, for a ``star-forming'' cell, given its gas density, $S$, and a time-interval,  we can use Eq. \ref{eq:P_fit} to model the SN-modified pressure, provided that the above conditions are satisfied. Note that we do not take into account the ``turbulence pressure'' here. As we will show in the next subsection, the Mach number is close to unity, which means that the pressure from the random motions is similar to that of the thermal pressure \citep{joung09}. In practice, a factor of 2 can be added to Eq. \ref{eq:P_fit} to make it the total pressure. Finally, an interesting coincidence is that the box size and duration of the simulations are roughly on the same orders of the cell size ($\sim$ 100 pc) and timestep ($\sim$ 10 Myr) of cosmological simulations, respectively.

\subsubsection{Other output parameters}

Table \ref{t:simIII} lists the input/output parameters for all simulations in this section. For each ISM phase, we show the mass fraction $f_m$, volume fraction $f_V$ and volume-weighted Mach number $\mathcal{M}_V$. Again, the numbers are obtained by averaging over 20 snapshots evenly sampled during the last $20\%$ of the simulation time.
The Mach number is defined as 
\begin{equation}
\mathcal{M} \equiv v/c_s, 
\end{equation}
where $v$ and $c_s$ are the local velocity (in the simulation box frame) and the local sound speed, respectively, and 
\begin{equation}
c_s \equiv \sqrt{\gamma k_B T/ \mu}, 
\end{equation}
where the adiabatic index $\gamma = 5/3$, and the molecular weight $\mu = 1.22,\ 0.588m_H$ for $T <10^4$, $> 10^4$ K gas, respectively.

The order of unity value of $\mathcal{M}$ may not be surprising, since the blast waves induced by SN explosions would dissipate until they decay to sound waves. Intriguingly, however, the deviation of $\mathcal{M}$ is fairly small for the three phases, despite the large parameter space $(\bar{n},S)$ we have explored. We find $\mathcal{M}_V =\ 0.5\pm 0.2,\ 1.2\pm 0.3,\ 2.3\pm 0.9$ for the hot, warm and cold phase, respectively. The corresponding mass-weighted $\mathcal{M}_m$ (not listed in the table) are $0.5\pm 0.3,\  1.0\pm 0.2,\ 2.6\pm 1.2$, respectively. The warmer phases have consistently lower $\mathcal{M}$. The Mach numbers of the cold phase lie within the observational range $\mathcal{M}\sim 1-4$, as derived from the thermal pressure distribution in the solar neighbourhood \citep{jenkins11}. However, we should caution that the cold clouds are usually resolved by $\lesssim 10$ grids in radius, and the velocity dispersion of the gas might be suppressed due to the limited numerical resolution. Finally, for a given density, $\mathcal{M}_V$ for the warm/cold phase seems to correlate (very) modestly with the $S$ for either $\bar{n}\geqslant 10 \cm-3$ or $\bar{n}\leqslant 0.1 \cm-3$. Note that for the former $\bar{n}$ range, the ISM is in a steady state (or metastable), and for the latter, thermal runaway (or metastable). Observationally, the velocity dispersion of H$\alpha$ may have a mild correlation with star formation rate \citep[e.g.][]{green14,arribas14}. Whether the velocity dispersion of HI has such a correlation is not clear; evidence for either side has appeared in literature \citep{tamburro09, rogers13,stilp13}, but the correlation is modest at most, if any. Our results are broadly consistent with the observations. However, here we are considering only SN as the turbulence driver; other processes, such as gravitational and/or magneto-rotational instability, can also play roles in driving the turbulence. Also note that for runs with larger $\bar{n}$, we use higher resolutions and smaller boxes. Thus the finer structures are increasingly resolved, especially for the dense phase. For the lower $\bar{n}$ runs, the resolution of the inner structure of the cold clumps may be limited by the finite cell size.

In Table \ref{t:simIII} we have also shown the fraction of the kinetic energy out of the total energy in the simulation domain, $f_{\rm{Ek}}$. This is obtained by averaging over 30 snapshots for the last 3\% of the simulation time. $f_{\rm{Ek}}$  is in the range of 15\% - 50\%, and the value is smaller for lower $\bar{n}$. This is consistent with that the denser media cool more efficiently, thus leaving a larger fraction of the energy in the kinetic form.

We define a ``heating time scale'' as
\begin{equation}
t_{\rm{heat}} \equiv \frac{e_{\rm{tot,V}}}{\dot{e}_{\rm{SN}}} = \frac{e_{\rm{tot,V}}}{S\cdot E_{\rm{SN}}},
\label{eq:t_heat}
\end{equation}
where $e_{\rm{tot}}$ is the volume-weighted energy density of the ISM at the end of the simulation, $\dot{e}_{\rm{SN}}$ is the time-averaged SN heating rate, and $E_{\rm{SN}} = 10^{51}$ erg. If the ISM reaches the thermal equilibrium, $\dot{e}_{\rm{SN}}$ is equal to the time-averaged cooling rate $\dot{e}_{\rm{cool}}$ (the other heating source, PEH, is an order of magnitude weaker than the SN overall), so $t_{\rm{heat}}$ is approximately the cooling timescale. If the ISM is in the thermal runaway state, $t_{\rm{heat}}/t_{\rm{sim}} \sim 1- \dot{e}_{\rm{cool}}/\dot{e}_{\rm{SN}}$, where $t_{\rm{sim}} $ is the duration of the simulation. From the table, $t_{\rm{heat}}$ is generally smaller than 1 Myr, much smaller than $t_{\rm{sim}}$, indicating that cooling is efficient in the ISM. Even in a thermal runaway state, $\dot{e}_{\rm{cool}}/\dot{e}_{\rm{SN}}$ is very close to unity. For extreme cases, such as $\bar{n}0035\_ S100$, $t_{\rm{heat}}/t_{\rm{sim}} \approx 0.2$, which implies that on average 20\% of the energy from each SN is retained in the ISM, instead of being radiated away. For the majority of the thermal runaway media, the retained fraction of energy per SN is about 1-5\%.

\section{Discussion}

\subsection{Randomness of SN position on simulation box scale}
\label{sec:random}

In the simulations described in Section 4, we have kept the positions of SN random. In this section we discuss how good this assumption is. Type Ia SN, which are from old stellar populations, are not correlated with the gas properties, and are thus close to random. Core collapse SN are more complicated. They descend from OB stars, and are mostly associated with star-forming clusters, which have abundant dense molecular clouds. Recent work have shown that the SN position relative to the dense clumps has a significant impact on the feedback. \cite{iffrig15} found that a SN is most disruptive to a molecular cloud when it explodes inside; a SN outside has only moderate effect on the cloud, but the feedback influence is felt much further in the low-density inter-cloud medium. Considering multiple SN, \cite{gatto15} discovered in a periodic box study that a random positioning leads to thermal runaway more easily than when the SN mostly explode within the clumps. \cite{hennebelle14} used a set-up with stratified medium and found that random SN lead to a higher gas density and pressure at high galactic latitude -- closer to what has been observed. The random positioning leaves most SN to explode in a rarefied medium, which enables an effective feedback on larger scales. 

To assess the SN position relative to the dense clumps, we emphasize that the non-negligible velocities of the OB stars should be taken into account. We note that some other effects can also lead core collapse SN to explode in a low-density environment. For example, molecular clouds disperse on a timescale of about 20 Myr, shorter than the life time of some SN progenitors, so even those in situ may explode in a rarefied gas; recent works \citep[e.g.][]{Bressert12,oey13} suggested that some OB stars are indeed born in the field, rather than in crowded clusters. But here we simply focus on the effect of the OB velocities. In particular, it would be very interesting to know how far OB stars can migrate before they explode as SN. To know the answer helps other important questions: (1) What fraction of core collapse SN explode inside molecular clouds versus outside? (2) Is the random positioning of SN assumed in our simulations a good approximation?

To calculate the displacement distribution of core collapse SN progenitors, we do the following simple Monte Carlo simulation: We adopt the distribution of space velocities of OB stars compiled by \cite{stone91}, which is a sum of two Maxwellians with $\sigma$ = 7.7 km/s and 28.2 km/s, respectively, and 46\% of O stars and 10\% of B stars belong to the high-velocity group. More recent work by \cite{dewit05} confirmed that roughly half of the O stars are ``runaways''. Each core collapse SN progenitor is randomly assigned a mass in the range $8\msun <  M_* < 50\msun $ according to the initial mass function $dN/dM_* \propto M_* ^{-2.3}$ \citep{kroupa01, chabrier03}, a speed according to the distribution mentioned above (if $M_*>16 \msun$, the progenitor is classified as an O star, otherwise a B star), and a random velocity direction. The life time of the progenitor is related to its mass by $t_{\rm{life}} = 10^{10}\yr (M_*/\msun)^{-2.5}$.

\begin{sidewaysfigure*}
\begin{center}
\begin{tabular}{c c}
\includegraphics[width=0.50\textwidth]{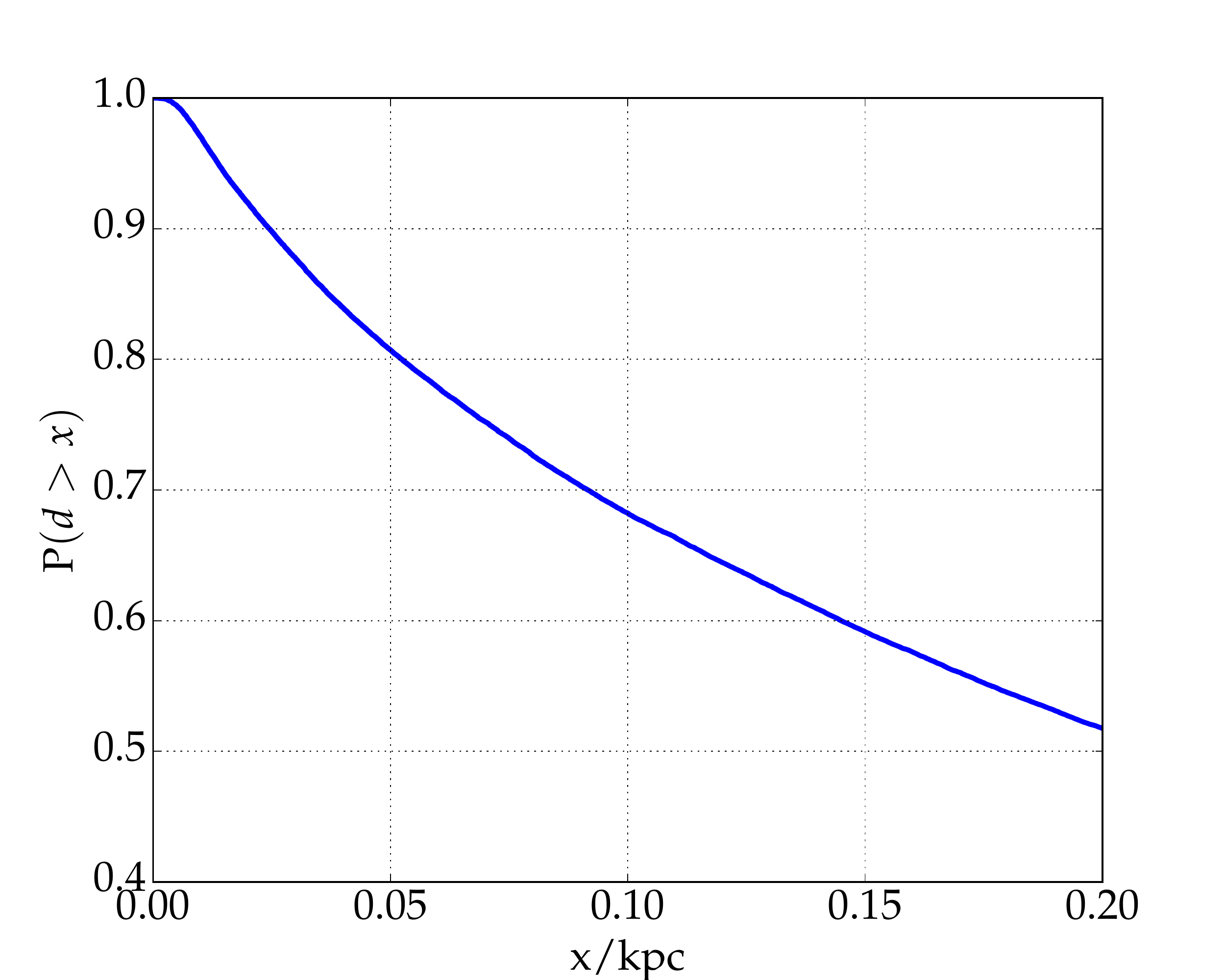} & \includegraphics[width=0.50\textwidth]{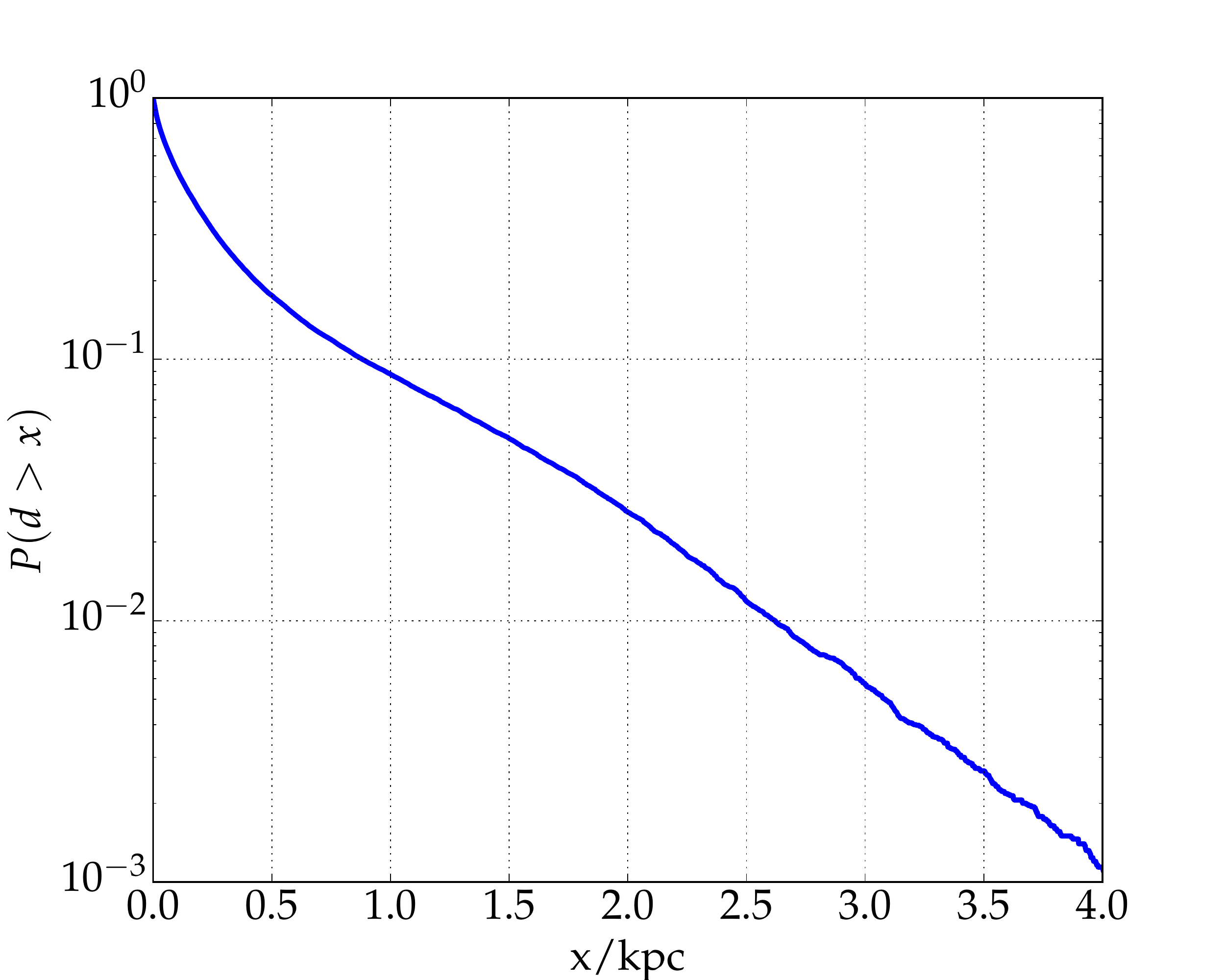}

\end{tabular}
\caption{ Left: Cumulative distribution of 3D displacement of core collapse SN, relative to the birth places of their progenitors, due to the velocities of OB stars. Roughly 68\% of core collapse SN have migrated more than 100 pc. Right: the projected 1D displacement. Nearly 10\% of OB stars can migrate $\gtrsim 1\kpc$. See Section \ref{sec:para} for the model description.  }
\label{f:runaway_3d_1d}
\end{center}
\end{sidewaysfigure*}

The displacement distribution is shown in Fig. \ref{f:runaway_3d_1d}. In 3D space (left panel), about 96\%, 82\%, 70\% of SN progenitors migrate away from their birth sites by distances greater than $10, 50, 100 \pc$, respectively. The giant molecular clouds appear filamentary \citep[e.g.][]{molinari10}, and the size is on the order of 100 pc in the long dimension and an order of magnitude smaller on the shorter ones. Our result suggests that roughly $\lesssim$ 5\% of SN would explode within these clouds. Those SN can cause the clouds to disperse and thus suppress star formation. On larger scales, the external gravitational field from the dark matter halo and stellar disk may preferentially confine the motion of the progenitor stars to a certain direction, so it makes more sense to calculate the projected 1D displacement distribution (right panel of Fig. \ref{f:runaway_3d_1d}). Approximately 17\%, 8.4\%, 2.5\% of the SN progenitors have displacements greater than 0.5, 1, 2 kpc, respectively. A significant fraction of runaway OBs are therefore likely to migrate out of the star formation regions, and explode in the low-density part of a galaxy, such as halo and inter-spiral arm regions, and may drive a wind there.

The sizes of our simulation domain range from a few tens to a few hundred parsec, and we want to know how the runaway stars affect the randomness of the SN position on this scale. We conduct the following comparison experiments: 
we run two additional simulations to $\bar{n}1\_ S200$ (``Random'') -- 
(1) the probability of where a SN progenitor is born is proportional to local gas density $\rho ^{1.5}$, and each SN is randomly assigned a displacement according to the displacement distribution in the left panel of Fig. \ref{f:runaway_3d_1d} (``HD+runaway", where HD stands for high-density); (2) same as (1) but no displacement, that is, SN explode where their progenitors are born (``HD"). All other input parameters, such as $\bar{n}$, $S$, the box size and resolution, remain the same. 

\begin{sidewaysfigure}[ht]
\begin{center} 
\includegraphics[width=1.2\textwidth]{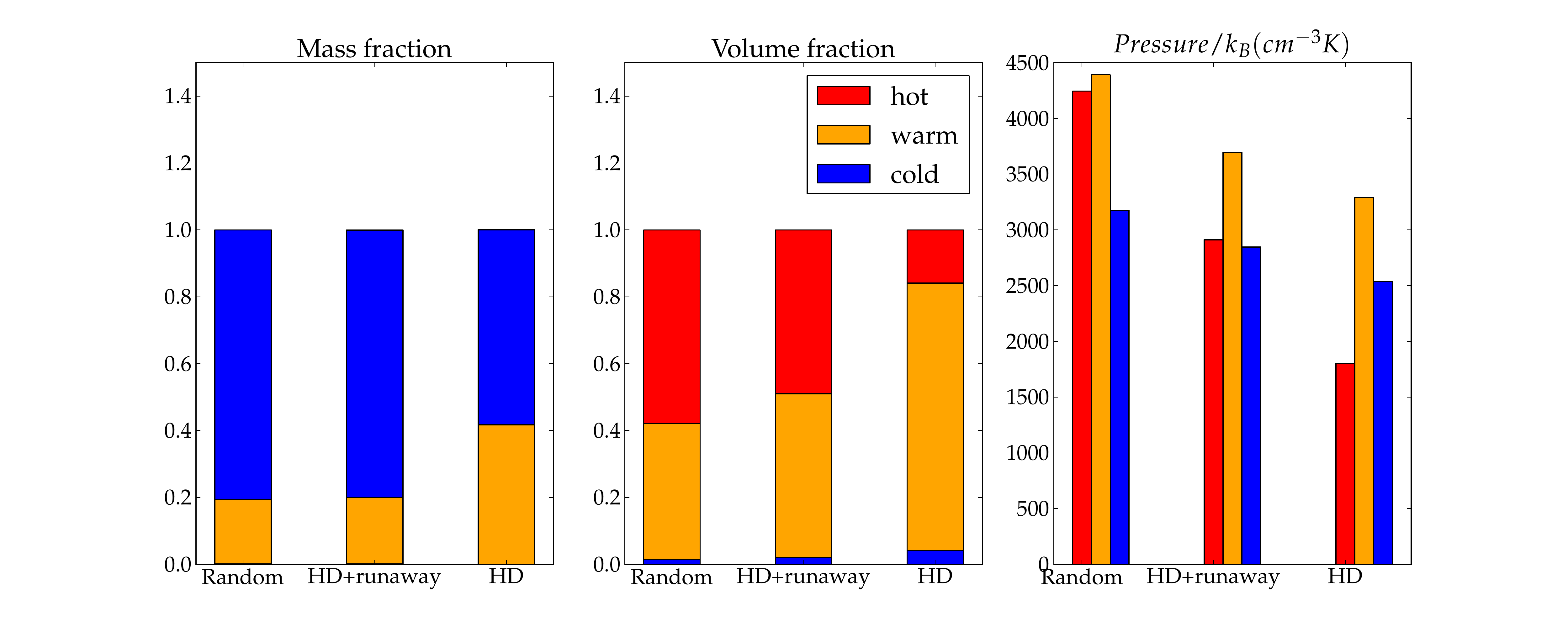}
\caption{Effects of SN location on the resultant ISM for $\bar{n}1\_ S200$ (box size 144 pc). Left to right: mass fraction, volume fraction, and volume-weighted pressure of the three ISM phases. ``Random": SN explode at random positions. ``HD": SN explode only in high-density clumps. ``HD + runaway": SN have a displacement from the high-density clumps, due to the velocities of their progenitors. The ISM from the ``Random'' and ``HD+runaway'' are similar, whereas that from ``HD'' has a much smaller $f_{\rm{V,hot}}$ and more warm gas. 
}
\label{f:ran_peak}
\end{center}
\end{sidewaysfigure}

The results are shown in Fig. \ref{f:ran_peak}: the properties of the ISM produced by ``HD+runaway'' fall in-between those in the other two cases.  In particular, displaced SN yield very similar volume/mass of each gas phase to random SN. The ``HD'' run produce more warm gas in mass and a much smaller volume fraction of the hot medium. This is understandable since the density peaks in the ISM consist mostly of the cold medium, and the energy released by a SN converts the cold medium to warm. The strong cooling in those dense clouds restrains the power of SN ($R_{\rm{cool}}\propto n^{-0.42}$), so they cannot impact the ISM on larger scales. This explains the smaller hot gas fraction and lower pressure. Those results are qualitatively consistent with \cite{gatto15}.  We conclude that based on the observed OB star velocities, the positions of core collapse SN are \textit{nearly} random on scales $\lesssim 150 \pc$.

\subsection{Connecting simulation data to real galaxy environments}
We have plotted $(\bar{n},S)$ encountered in the MW disk to halo on Fig. \ref{f:fvh} (blue solid line), as described in Section \ref{sec:para}. For the disk average $\bar{n}1\_ S200$, the ISM is in the ``transitional stripe". Of course, the disk average does not take into account the clustering effect of the SN. OB stars are born mainly in OB associations, and eventually 60\% of core collapse SN appear ``grouped''. Since the disk average $S$ makes the ISM in the transitional region, it is not surprising that any grouping in space and/or time can lead to thermal runaway, assuming the same gas mean density. Indeed, if SN explode within a sphere of $R\sim$ 100 pc every 0.3 Myr as for a super-bubble \citep{maclow88}, which corresponds to $S\sim 800 \kpc^{-3}\myr-1$, it is well within the thermal runaway regime for $\bar{n}=1\cm-3$ (Fig. \ref{f:fvh}).

As one goes up to the halo, the mean density is smaller, and the SN rate is such that $f_{\rm{V,hot}}$ is at first larger and the ISM goes into the thermal runaway region. This suggests that at least part of the halo is not in hydrostatic equilibrium. Above a certain height, $f_{\rm{V,hot}}$ drops again, which may suggest that the gaseous halo is convective, but a conclusion cannot be made without a simulation with stratified medium and an external gravitational field \citep{joung06, gent13,creasey13}. 

We have also plotted the scaling relation of $S$ and $\bar{n}$ as suggested by the Schmidt star formation relation $S \propto \bar{n}^{1.5}$ (green dashed line). The relation is normalized such that the line passes through the MW average $\bar{n}1\_ S200$. For data points that lie around this relation, $f_{\rm{V,hot}}$ is smaller for higher densities. For $\bar{n}=30 \cm-3$, the ISM around the Schmidt relation has a \fvh of only a few percent. Surprisingly, for ISM in star-burst regimes, little hot gas exists and the medium is thermally stable. Observationally, however, powerful winds are ubiquitously associated with such high star formation rates, and the mass loading ($\equiv \dot{M}_{\rm{outflow}}/ \dot{M}_{\rm{SF}}$) is on the order of unity or even higher \citep[e.g.][]{heckman00,steidel04}. Note that we have found the PEH rate to be a very important factor in determining the thermal state of the ISM. If the PEH rates we have adopted are reasonable, then the data seem to suggest that one needs other mechanism(s) to drive a wind for the high density regions. For example, ``clustering'' of SN may work, although a factor of 10 or more increase in $S$ is needed for $\bar{n}\gtrsim 3 \cm-3$ above the average value to have a thermally-driven wind. Another factor is the pre-SN feedback, such as photo-ionization, radiation pressure and winds from massive stars, which can create low density tunnels, facilitating SN energy to leak out \cite[e.g.][]{rogers13}. Alternatively, runaway OB stars, which can easily migrate more than a hundred parsecs, may lead to a significant fraction of core collapse SN exploding outside the immediate high density SF clouds. For a low density medium $\bar{n} \lesssim 1\cm-3$, the critical SN rate for the ISM to have a thermal runaway is much less stringent. Another possible mechanism for wind launching is by the cosmic rays. As the cosmic rays diffuse out, the pressure gradient exerts accelerating force on the baryonic gas. Recent simulations have shown that this mechanism is promising to drive winds with a reasonable mass loading \citep{ uhlig12, hanasz13,salem14}.

\subsection{Periodic box}

In the experiments where multiple SN shape the ISM, we have applied periodic boundary condition and initially uniform medium. We briefly discuss their applicability here. 

In a periodic box, the total mass is fixed, and the gas is not allowed to expand beyond the simulation domain. This means that such a box is good at capturing processes that have scales smaller than its size, which do not result in net ingoing/outgoing flux on the outer boundary. For our experiments, this is the case for those reaching steady states. For the thermal runaway cases, the pressure keeps rising and the box is accumulating energy. In real galactic environment, the hot gas is subject to expansion if the ISM pressure around is not sufficient to stop it. The expansion can lower the pressure and $f_{\rm{V,hot}}$ \citep{deavillez04}. Such an evolution thus cannot be faithfully captured by a periodic box.

On the other hand, we want to emphasize that the onset criterion for the thermal runaway, i.e. when the SNRs overlap, as given in $S_{\rm{cirt}} (\bar{n})$, should not depend on whether the box is open or periodic -- it only depends on the local density of the gas. Furthermore, even initially the outgoing flow is in an explosive state, it may adjust so that each patch of the flow comes to an equilibrium state. The length scale of the galactic winds is $\sim$1-10 kpc, much larger than $R_{\rm{cool}}$ of a single SN, and the dynamical time scale is $\sim$ Gyr, much longer than what takes to reach a steady state. So when the gas on large scale is in a steady state, the effect of SN may be regarded as ``microphysics'' and incorporated in the effective equation of state in a cosmological simulation.

A related issue is the density distribution in the box. In our simulations we keep gas uniform as the initial condition. We note that in real galaxies, especially disk galaxies, the ISM is stratified in the vertical direction. Therefore, the uniform assumption only holds when the processes we study have scales smaller than the scale height of the gas. The size of our simulation domain, which is $6R_{\rm{cool}}$, is generally smaller than the gas scale height in the MW, given a mean ISM density. For example, the thickness of the neutral gas layer is about 180 pc with a mean density $\sim 1\cm-3$, the warm ionized medium has a scale height of about $1830\pc$ with the density $0.014 \cm-3$ \citep{gaensler08}. For comparison, the box size is $142.2\pc$ for $\bar{n}=1\cm-3$, and $585.6\pc$ for $\bar{n}=0.035 \cm-3$. So the non-stratified medium is probably not a bad assumption. That said, the gas scale height in extragalactic systems are usually not well-constrained, and SN themselves may play an important role in determining it \citep[e.g.][]{shetty12}.

In disk galaxies, clustered SNR in a thermal runaway state can form ``super-bubbles'' that may break out of the disk and transfer materials into the halo \citep{tomisaka86, maclow88}. The outflowing gas forms galactic winds and/or fountains that regulate the circum-galactic environment \citep{heckman00, steidel04}. These can only be captured in a larger box with more realistic set-up. For example, the medium is stratified, which has outflow outer boundary conditions in the vertical direction and external gravitational field \citep[e.g.][]{joung06, creasey13}. We will postpone the study of SN feedback in such cases to future work.

\subsection{Thermal conduction and magnetic fields}

Thermal conduction can be an important process in a multiphase medium to drive mass and energy transfer among different phases. In the model of \cite{mckee77}, evaporation through thermal conduction plays a major role in producing the warm phase from the dense, cold clumps in the post-shock region. Very small clouds bathed in a hot medium can be quickly evaporated. As mentioned earlier, thermal conduction is not included in our simulation. The warm phase is maintained mainly through the balance of the shock heating/PEH and cooling. We did some calculations to estimate the role of conduction for our simulations. For the multiphase ISM models we adopt in Section \ref{sec:II}, conduction actually causes hot gas to condense on the clouds \citep{cowie77}, and the evaporation is only important when very close to the explosion center, where the temperature is high and clouds are shock-stripped to small cold cores. The difference between our simulation and \cite{mckee07} arises from that our clouds, which are created and destroyed by SN interactions, are larger by a factor of a few than what was adopted in the \cite{mckee07} model, which was inspired by the HI observation at that time. Smaller clouds would result in a more effective mixing between the different phases in a number of ways, e.g. Kelvin-Hemholtz instability, and photo-evaporation by massive stars \citep{mckee84}. To fully evaluate the role of conduction, one also has to consider the magnetic field, which can cause significant anisotropy. Conduction is practically inhibited perpendicular to the field lines. When a blast wave travels in a HDMM, it tends to wrap the field lines around the clouds \citep{semenov80,dursi08}, which suggests that the conduction between clouds and the hot medium may not be very important. The actual impact of conduction considering magnetic field is not clear and is definitely worth further study. 

The magnetic field itself can also be important dynamically. Local observations show that the magnetic pressure is a few times larger than gas thermal pressure \citep{heiles05}. \cite{slavin92} has found that in a uniform medium, the magnetic pressure at late stages of the SNR can thicken the shell and thus make the hot bubble smaller. This may change the critical SN rate for thermal runway at a given density. A MHD simulation is needed to quantify the role of the magnetic field in shaping the ISM \citep[e.g.][]{shin08,hill12}.

\section{Conclusion}
In this paper we explore the interplay between SN and the three-phase ISM. We conduct three sets of experiments to compare the evolution of a SNR in a uniform and a hot-dominant multiphase medium, and to allow SN to self-consistently shape the ISM under various environments.
The main results are summarized as follows:

1. For a uniform medium, the evolution of a SNR has four well-defined stages after the initial free expansion: Sedov-Talor, (transient) ``pressure-driven snowplow", momentum-conserving and close-up. Thermal energy drops precipitously once the shock becomes radiative as $E_{\rm{th}} \propto R_p ^{-6\sim -8}$, while kinetic energy drops later and more slowly as $E_k \propto R_p ^{-3}$. The $R_p - t$ relation and the energy/momentum feedback vs $R_p$ can be well described by piece-wise power-law functions. The power-law indices vary little over a wide range of gas densities and photoelectric heating rates (Table 1).

2. In a hot-dominated multiphase medium, a SNR evolves very differently from the uniform case or a two-phase medium. There are no distinct stages of evolution. The blast wave initially interacts with the clouds, but later travels much faster and mainly in between the clouds (Eq. \ref{eq:r_sh}). 
The SNR thus sweeps up much less mass, therefore developing less momentum at a given $R_p$ (Fig. \ref{f:E_t_allmulti}). A radiative-efficient thin shell never forms at the blast wave front, and the overall cooling rate is suppressed. The hot phase helps to retain the SN energy and facilitates energy transfer to much larger scales.  The spatial-temporal domain a SN is enlarged by a factor of $>10^{2.5}$ in a HDMM.

3. We study the ISM self-consistently shaped by multiple SN, covering a wide range of SN frequency $S$ and gas average density $\bar{n}$. We find that the classic view of the three-phase ISM shaped by SN, as envisioned by \cite{mckee77}, needs to be extended, depending on $\bar{n}$ and $S$. The main differences we have found are the following: First, a certain phase can be absent, or exist only very briefly. Second, equilibrium solutions may not exist when SN explode at a sufficiently high rate, and the ISM undergoes thermal runaway. Third, the pressure of different phases may not be in equilibrium; the hot phase can be over- or under-pressured. 

4. We find that $f_{\rm{V,hot}}\sim 0.6 \pm 0.1$ marks the transition of ISM from a steady state to thermal runaway, along with the change of hot gas topology. When $f_{\rm{V,hot}}\lesssim 0.6 \pm 0.1$, SNRs evolve as individual bubbles; if $f_{\rm{V,hot}}$ can reach $0.6 \pm 0.1$, the hot bubbles connect to each other. The connecting topology makes any new SNR have a larger spatial-temporal domain of influence, which reinforces the overlapping of the blast waves. As a result, the overall heating dominates over cooling and the ISM undergoes thermal runaway. An ISM with $f_{\rm{V,hot}}\sim 0.5-0.7$ is in a metastable state, and may transition into thermal runaway.

5. The PEH has a surprisingly strong impact on $f_{\rm{V,hot}}$. For a fixed $(\bar{n},S)$, $f_{\rm{V,hot}}$ decreases as the PEH rate increases. In particular, for $\bar{n}\gtrsim 3 \cm-3$, assuming that the PEH is proportional to $S$, $f_{\rm{V,hot}}$ is confined to $\lesssim 0.6$ for a wide range of $S$, so that the medium is stable and does not form a wind (Fig. \ref{f:fvh}). The critical SN rates for the onset of the thermal runaway is roughly 
$S_{\rm{crit}} =  200 (\bar{n}/1\cm-3)^k (E_{\rm{SN}}/10^{51}\erg)^{-1} \kpc^{-3} \myr-1$, where $k = (1.2,2.7)$ for $\bar{n} \le 1$ and $>1 \cm-3$, respectively. 

6. The pressure of the ISM relates to $f_{\rm{V,hot}}$, mean gas density $\bar{n}$ and photoelectric heating rate $\Gamma_{\rm{pe}}$ through Eq. \ref{eq:neq}-\ref{eq:P_pred}. The results do not depend on whether the ISM is in a steady state or thermal runaway. A simple fitting formula $P_V(\bar{n},S,t)$ can be used as an effective equation of state considering SN feedback:
\begin{equation}
P_V/k_B= \begin{cases}
2.9\times 10^3 \cm-3\rm{K} (\frac{S}{100 \kpc^{-3} \myr-1 })^{0.87} (\frac{\bar{n}}{1\cm-3})^{-0.33}\  &\text{if $S<S_{\rm{crit}}$, } \\
9.6\times 10^3 \cm-3\rm{K} (\frac{S}{1000 \kpc^{-3} \myr-1 })^{1.87} (\frac{\bar{n}}{1\cm-3})^{-1.59} (\frac{t}{10\rm{Myr}})\  &\text{if $S\geqslant S_{\rm{crit}}$. } \nonumber
\end{cases}
\end{equation}

7. The local Mach numbers $\mathcal{M}$ of the three ISM phases show surprisingly small variations, despite the 5 orders of magnitude span of $(\bar{n},S)$. We find that $\mathcal{M}_V \approx \ 0.5\pm 0.2, \ 1.2\pm 0.3,\ 2.3\pm 0.9$ for the hot, warm and cold phase, respectively.

8. We calculate the displacement distribution of the core collapse SN from the observed velocities of OB stars, which shows that on scales $\lesssim 150\pc$, SN explode almost at random positions, and nearly 10\% of OB stars can migrate $\gtrsim 1\kpc$ (Fig. \ref{f:runaway_3d_1d}). The latter runaway OBs have a chance to migrate into low-density regions where it is easier to drive a wind.

\section*{Acknowledgement} 
We are grateful to the anonymous referee, who gave comments that has greatly improved the clarity of the draft. We thank Bruce Draine, Andrea Gatto and David Schiminovich for useful discussions and Britton Smith for the help with Grackle. Roger Chevalier, Mordecai-Mark Mac Low, Christopher McKee and Eve Ostriker kindly read the manuscript and gave many helpful comments. The computing resources are provided by the Hecate cluster of Princeton University. The research is supported by NASA grant NNX11AI23G, NNX12AH1G, NSF grants AST-1210890 and AST-1312888. This work was partly performed at the Aspen Center for Physics, which is supported by NSF grant PHY-1066293.

\newpage

\end{document}